\def\preprint{1}			

\ifdefined\preprint
\documentclass[preprint,review,12pt]{elsarticle}
\fi
\ifdefined\wordcount
\documentclass[final,3p,times,twocolumn]{elsarticle}
\fi
\ifdefined\final
\documentclass[final,3p,times,twocolumn]{elsarticle}
\fi

\DeclareMathSizes{7}{7}{6}{6}

\usepackage{graphicx,stfloats}
\usepackage{color}

\usepackage[fleqn]{amsmath} 
\usepackage{hyperref} 
\usepackage{soul}

\usepackage{tabularx}
\newcolumntype{Y}{>{\centering\arraybackslash}X}
\usepackage{cleveref}
\usepackage{tikz}
\usepackage{natbib}
\usepackage[separate-uncertainty = true]{siunitx}
\usepackage{nicefrac}
\usepackage{caption}
\usepackage{subcaption}
\usepackage{adjustbox}
 \usepackage{multirow}
\usepackage{wasysym}
\usepackage{gensymb}
\usepackage{amsthm}

\newcommand\chiLL{\chi_{\Lambda\Lambda}}
\newcommand\chiLLref{\chiLL^{\rm ref}}
\newcommand\chiLLnot{\chiLL^{\rm 0}}
\newcommand\ammonia{\rm NH_{3}}
\newcommand\methane{\rm CH_{4}}
\newcommand\hydrogen{\rm H_{2}}
\newcommand\nitrogen{\rm N_{2}}
\newcommand\NO{\rm NO}

\newcommand\NtwoO{\rm N_{2}O}
\newcommand\NOx{{\rm NO}_{x}}
\newcommand\COtwo{{\rm CO_{2}}}
\newcommand\CO{{\rm CO}}
\newcommand\LambdaAst{\Lambda^{\!*}}
\newcommand\Ntwo{\rm N_{2}}
\newcommand\Zstoi{Z_{\rm st}}

\biboptions{sort&compress}

\journal{Combustion and Flame}

\begin{document}

\begin{frontmatter}

\title{Rush-to-equilibrium concept for minimizing reactive nitrogen emissions in ammonia combustion}

\author[fir]{Hernando Maldonado Colmán\corref{cor1}}
\ead{hm2524@princeton.edu}

\author[fir]{Michael E. Mueller}

\address[fir]{Department of Mechanical and Aerospace Engineering, Princeton University, Princeton, NJ 08544, USA}
\cortext[cor1]{Corresponding author.}

\begin{abstract}
	
\par
Ammonia is a zero-carbon fuel that has been receiving increasing attention for power generation and even transportation. Compared to hydrogen, ammonia's volumetric energy density is higher, is not as explosive, and has well established transport and storage technologies. However, ammonia has poor flammability and flame stability characteristics and more reactive nitrogen emissions (nitrogen oxides, nitrous oxide) than hydrocarbon fuels, at least with traditional combustion processes. Partially cracking ammonia addresses its flammability and stability issues, through on-board catalysts or autothermal crackers, into a mixture of ammonia, hydrogen, and nitrogen. However, reactive nitrogen emissions remain a challenge, and mechanisms of their emissions are fundamentally different in ammonia and hydrocarbon combustion. While rich-quench-lean ammonia combustion strategies have shown promise, the largest contributions to reactive nitrogen emissions are the unrelaxed emissions in the fuel-rich first stage due to overshoot of thermodynamic equilibrium within the reaction zone of premixed flames coupled with finite residence times available for relaxation to equilibrium. This work introduces a rush-to-equilibrium concept for partially cracked ammonia combustion, which aims to reduce the unrelaxed reactive nitrogen emissions in finite residence times by accelerating the approach to equilibrium. In the concept, a flow particle is subjected to a decaying mixing rate as it transits the premixed flame. This plays an important role in mitigating the mixing effects that prevents the flow particle approach to equilibrium, and promoting the chemistry effects to push the particle toward equilibrium, all while considering residence times typical of gas turbines. Evaluated with a state-of-the-art combustion model at gas turbine conditions, the concept shows the potential for a reduction in reactive nitrogen emissions by an order of magnitude at even modest mixing rate decay rates. It is also shown that the concept works irrespective of cracking extent, pressure, temperature, etc. A brief discussion of possible practical implementation reveals reasonable geometric and flow parameters characteristic of modern gas turbine combustors.

\noindent\textbf{Novelty and Significance Statement}

A novel rush-to-equilibrium combustion concept is proposed with the aim of reducing reactive nitrogen emissions, which include nitrogen oxides and nitrous oxide, from partially cracked ammonia combustion at gas turbine conditions. Reactive nitrogen emissions are elevated in partially cracked ammonia combustion systems because insufficient residence time is available to reach thermodynamic equilibrium. A concept is proposed to address this issue, leveraging decaying mixing rates, without modifying typical gas turbine residence times by accelerating the approach to thermodynamic equilibrium. The new concept is demonstrated of being capable of significantly reducing reactive nitrogen emissions (by a factor of an order of magnitude). Finally, implementation of the new concept is showed to be practically feasible.

\end{abstract}


\end{frontmatter}

\textbf{Authors contribution}

\underline{HMC}: performed research, software development, writing -- original draft \& editing

\underline{MEM}: designed research, writing -- review \& editing, project administration


\ifdefined \wordcount
\clearpage
\fi

\section{Introduction}
\label{sec:introduction}

Decarbonizing combustion energy conversion has become a critical endeavor to combat climate change, and zero-carbon fuels will play a vital role in achieving net-zero carbon emissions in the global economy as quickly as possible. While hydrogen has received the most attention as a future zero-carbon fuel, ammonia is garnering increasing attention due to its ease of transport compared to hydrogen and well established infrastructure due to its use in agriculture~\cite{valera2018ammonia,kobayashi2019ammonia,valera2021review,elbaz2022review}. However, challenges arise in burning ammonia due to its low flammability, its poor flame stability, and its emission of reactive nitrogen species, including nitrogen oxides ($\NOx$) and nitrous oxides ($\NtwoO$)~\cite{kobayashi2019ammonia}. Improvements to the combustibility of ammonia can be addressed by partially cracking ammonia, achieved through on-board catalysts or autothermal crackers \cite{monge2024ammonia}, resulting in a mixture of ammonia, hydrogen, and nitrogen (AHN), which can be tailored to have similar combustion properties to methane~\cite{jiang2020updated,wiseman2021comparison}. The grand challenge then is to develop ammonia combustion processes with low reactive nitrogen emissions, which is critical to its overall environmental benefit or potential harm~\cite{bertagni2023ammonia}.

Recent work in the literature suggest several strategies have been investigated to improve ammonia combustion, namely swirl stabilized combustion \cite{franco2021characteristics,mashruk2022nitrogen,mashruk2022evolution,pugh2021investigation,mashruk2023chemiluminescent},  moderate or intense low oxygen dilution (MILD) combustion \cite{sorrentino2019low,ferrarotti2020influence,rocha2021combustion,mohammadpour2022reaction}, and staged combustion approach in a rich-quench-lean (RQL) configuration~\cite{somarathne2017numerical,okafor2019towards,wu2020no,bovzo2021humidified,mashruk2021rich,okafor2021liquid,li2022effects,heggset2024numerical}. The latter suggests that burning ammonia using the RQL configuration can mitigate reactive nitrogen emissions \cite{ditaranto2021experimental}. Indeed, fuel-lean ammonia combustion produces very high nitrogen oxide emissions compared to fuel-rich ammonia combustion~\cite{somarathne2017numerical,kobayashi2019ammonia}. Therefore, a fuel-rich first stage burns much of the ammonia without excessive nitrogen oxide emissions, and a fuel-lean second stage burns the resulting hydrogen and any remaining ammonia. Considered differently, the first stage essentially cracks the ammonia to minimize the ammonia burned in the second stage. Ammonia combustion strategies have been developed based on hydrocarbon-fueled RQL concepts~\cite{guthe2008reheat,lefebvre2010gas,emerson2025nonpremixed} with quite low reactive nitrogen emissions~\cite{okafor2019towards,indlekofer2023numerical}. Another work \cite{rocha2021combustion} also suggests that MILD combustion approach also leads to low emissions that are comparable to the RQL configuration \cite{ferrarotti2020influence,rocha2021combustion,mohammadpour2022reaction,elbaz2022review}.

While ammonia combustion systems based on hydrocarbon-fueled system concepts have been somewhat successful in achieving relatively low reactive nitrogen emissions, ammonia and hydrocarbons have different chemical pathways to reactive nitrogen emissions, which fundamentally changes their nature. Recent studies \cite{gubbi2023air,gubbi2024evaluation,heggset2024numerical} indicate that burning ammonia in RQL gas turbine, with residence times typical of hydrocarbon gas turbines, results in significantly higher $\NOx$ emissions compared to thermodynamic equilibrium. This increase is attributed to the ``unrelaxed'' component of $\NOx$ emissions originating from the first stage, albeit still lower that a lean premixed single-stage combustor running on ammonia. These unrelaxed emissions result from the overshoot of reactive nitrogen emissions within the reaction zone of ammonia flames, well above their thermodynamic equilibrium values, and the excessive time required for these emissions to relax toward thermodynamic equilibrium. Unfortunately, practical constraints, such as the size and cost of the gas turbine, limit the residence time in the combustion chamber. Of course, with hydrocarbon fuels, long residence times are undesirable anyway since reactive nitrogen emissions rapidly increases as equilibrium is approached.

This work introduces a novel concept for ammonia (more generally, AHN) combustion, termed `rush-to-equilibrium', aiming to minimize unrelaxed reactive emissions without increasing residence times. The rush-to-equilibrium concept is analyzed with a premixed manifold combustion model \cite{mueller2020physically}, which is first introduced. Next, using the manifold model, characteristics of AHN combustion and methane combustion are compared, with specific emphasis on the interplay between mixing and chemistry in premixed flames, to motivate the rush-to-equilibrium concept. Then, the rush-to-equilibrium concept is introduced, leveraging a decaying mixing rate, and an in-depth analysis is undertaken to demonstrate the degree of reactive nitrogen emissions reduction that can be achieved. Finally, the practical feasibility and implementation of the rush-to-equilibrium combustion concept is then briefly discussed.

\section{Manifold model for premixed combustion}
\label{sec:maniL}

Combustion is governed by a high-dimensional thermochemical state including the temperature $T$ and species mass fractions $Y_{k}$, with $k=1,\dots,N_{s}$ and $N_{s}$ the number of species in the gas mixture. The dimensionality can be reduced by projecting the thermochemical state onto a reduced-order manifold~\cite{pope2013manifold}. For premixed combustion, the thermochemical state is projected onto a one-dimensional manifold in the progress variable $\Lambda$, which is zero in the unburned gases and unity at thermodynamic equilibrium.

Starting from the governing equations for the species mass fraction (and temperature), following the notation of Mueller~\cite{mueller2020physically}, the one-dimensional equilibrium manifold equations in progress variable for the species mass fractions are given by
\begin{equation}
    \frac{\partial Y_{k}}{\partial \Lambda}\dot{m}_{\Lambda}=\frac{\rho \chi_{\Lambda\Lambda}}{2}\frac{\partial^2 Y_{k}}{\partial \Lambda^2}+ \dot{m}_{k},
    \label{eq:Yk_transport}
\end{equation}
where $\dot{m}_\Lambda$ and $\dot{m}_k$ are the progress variable and species mass fraction source terms, respectively, $\rho$ is the density, and $\chiLL=2D\nabla\Lambda\cdot\nabla\Lambda$ is the progress variable dissipation rate, that is, the mixing rate, where $D$ is the thermal diffusivity. The progress variable $\Lambda$ is then defined as
\begin{equation}
    \Lambda=\frac{Y_{R}-Y_{R,\rm u}}{Y_{R,\rm eq}-Y_{R,\rm u}},
    \label{eq:Lambda}
\end{equation}
where the subscripts u and eq correspond to the unburned and equilibrium mixture. The subscript $R$ corresponds to the reference mass fraction $Y_{R}=\sum_{k\in R}a_{k}Y_{k}$, comprised of a weighted linear combination of product mass fractions. The corresponding progress variable source term is then defined as
\begin{equation}
    \dot{m}_{\Lambda}=\frac{1}{\partial Y_{R}/\partial \Lambda} \dot{m}_{R}\:.
    \label{eq:mdotL}
\end{equation} 
The reference species for AHN is $\rm H_{2}O$ and weight $a_{R}=1$; for methane, the reference species are $\COtwo$, $\CO$, $\rm H_{2}O$, and $\hydrogen$, with weights $a_{k\in R}=1$, 1, 1, and 10, respectively. In Eq.~\ref{eq:Yk_transport}, the basic balance of the premixed manifold equations is the progress variable as the ``arrow-of-time'' on the left-hand-side balance by mixing and chemistry effects on the right-hand-side.

Unstretched premixed flames can be analyzed to understand the relative roles of chemistry and mixing in the manifold equations. To model unstretch premixed flames in progress variable space, an additional transport equation for the gradient of the progress variable $g_{\Lambda}=\nabla \Lambda$ can be solved with the equilibrium manifold equations, similar to work of Scholtissek et al.~\cite{scholtissek2019self}. In this work, the equation is adapted to the progress variable $\Lambda$ \cite{mueller2020physically} as
\begin{equation}
 \dot{m}_\Lambda \frac{\partial g_{\Lambda}}{\partial \Lambda} = {g^{2}_\Lambda} \frac{\partial^2}{\partial \Lambda^2}(\rho D g_{\Lambda}) + g_{\Lambda} \frac{\partial \dot{m}_\Lambda}{\partial \Lambda},
    \label{eq:grad_L_eq}
\end{equation}
where the left-hand-side term corresponds to convective effects ($\Lambda$-drift), and the right-hand-side terms correspond to diffusive and chemical effects \cite{scholtissek2019self}. The manifold solution for unstretched flames will be useful to show later how the emissions behave in methane and (partially cracked) ammonia combustion at different compositions. Mixing effects in turbulent combustion lead to stretched flame configurations, and Eq.~\ref{eq:grad_L_eq} will no longer be representative of such cases. However, results are not sensitive to these profiles, and the $g_{\Lambda}$ transport equation is omitted, which makes computations more efficient for assessing the rush-to-equilibrium concept. To close the manifold equations for stretched premixed flames, the dependence of the progress variable dissipation rate on progress variable takes a presumed form \cite{mueller2020physically,novoselov2021two}:
\begin{equation}
    \begin{aligned}
    \chiLL (Z,\Lambda) &= \chiLLref \exp \left[-2\left(\operatorname{erfc}^{-1}\left(\frac{2 Z\left(1-\Zstoi \right)}{\Zstoi (1-Z)+Z\left(1-\Zstoi\right)}\right)\right)^2\right] \\
    & \times \exp \left[-2\left(\operatorname{erfc}^{-1}(2 \Lambda)\right)^2\right],
    \end{aligned}
\label{eq:chiLL_def}
\end{equation}
where $\chiLLref$ is a reference dissipation rate evaluated at the stoichiometric mixture ($\Zstoi$) and reference progress variable $\Lambda=0.5$, and $Z$ is the mixture fraction (based on the equivalence ratio of the unburned gas composition). While the manifold model can generally accommodate differential diffusion \cite{mueller2020physically,lacey2024data}, for convenience, a unity effective Lewis number assumption is made that does not impact the results presented subsequently in the manuscript.

The above manifold equations from Eq.~\ref{eq:Yk_transport} are an ``equilibrium'' manifold formulation~\cite{mueller2020physically} in which the local flame structure in progress variable instantaneously equilibrates to the instantaneous progress variable dissipation rate. As discussed below, the rush-to-equilibrium concept relies on an unsteady mixing rate, that is, an unsteady progress variable dissipation rate $\chiLLref\left(t\right)$, so, to ensure any flow history effects are considered, a ``non-equilibrium'' manifold formulation~\cite{mueller2020physically} is also considered that includes an additional unsteady history term on the left-hand-side:
\begin{equation}
    \rho\frac{\partial Y_k}{\partial t} + \frac{\partial Y_{k}}{\partial \Lambda}\dot{m}_{\Lambda}=\frac{\rho \chi_{\Lambda\Lambda}}{2}\frac{\partial^2 Y_{k}}{\partial \Lambda^2}+ \dot{m}_{k},
    \label{eq:Yk_transport_u}
\end{equation}
where, without loss of generality, this history term is indicated with a partial derivative in physical time (cf. Ref.~\cite{mueller2020physically}).

The corresponding progress variable transport equation is written as
\begin{equation}
    \rho\frac{\partial \Lambda}{\partial t} + \rho u_{j} \frac{\partial \Lambda}{\partial x_{j}}   = \frac{\partial}{\partial x_{j}}\left(\rho D \frac{\partial \Lambda}{\partial x_{j}}\right) + \dot{m}_{\Lambda},
    \label{eq:L_transp_eq}
\end{equation}
where $u_{j}$ correspond to the flow velocity. Given a time-varying mixing rate $\chiLLref\left(t\right)$, rather than look at the global flame structure in progress variable space at a particular time, a Lagrangian analysis will follow a reacting flow element that samples a specific temporal trajectory of progress variable. The progress variable of the reacting flow element $\LambdaAst\left(t\right)$ (distinct from progress variable space denoted by the simple $\Lambda$) is obtained by reorganizing the progress variable transport equation (Eq.~\ref{eq:L_transp_eq}) in a Lagrangian perspective:
\begin{equation}
    \rho\frac{d \LambdaAst}{d t} = \frac{\partial}{\partial x_{j}}\left(\rho D \frac{\partial \LambdaAst}{\partial x_{j}}\right) + \dot{m}_{\LambdaAst},
    \label{eq:maniL_mod}
\end{equation}
where $d/dt=\partial/\partial t+u_{j}\partial/\partial x_{j}$ is the Lagrangian derivative. To solve this equation with the non-equilibrium manifold equations, the diffusion term on the right-hand-side can be recast in progress variable space such that the evolution of $\Lambda$ can be determined from the global flame structure:
\begin{equation}
    \frac{\partial}{\partial x_{j}}\left(\rho D \frac{\partial \LambdaAst}{\partial x_{j}}\right) = \left.\left[ \frac{1}{4}\frac{\partial \rho\chiLL}{\partial \Lambda} + \frac{\chiLL}{4D}\frac{\partial\rho D}{\partial \Lambda}\right]\:\right|^{\LambdaAst},
    \label{eq:new_mixingterm}
\end{equation}
\noindent
so recasting Eq.~\ref{eq:maniL_mod} leads to
\begin{equation}
    \rho\frac{d \LambdaAst}{d t} = \left.\left[ \frac{1}{4}\frac{\partial \rho\chiLL}{\partial \Lambda} + \frac{\chiLL}{4D}\frac{\partial\rho D}{\partial \Lambda} + \dot{m}_{\Lambda}\right]\:\right|^{\LambdaAst}.
    \label{eq:maniL_mod2}
\end{equation}
In the right-hand-side of Eq.~\ref{eq:maniL_mod2}, $[\cdot]|^{\LambdaAst}$ indicates that it is evaluated at $\LambdaAst$ sampled from the solutions to Eq.~\ref{eq:Yk_transport_u}. Also, $\LambdaAst$ is used to sample the global flame structure as they both vary with Lagrangian time together. In other words, the evolution of $\LambdaAst$ is solved together with the ``non-equlibrium'' manifold equations in Eq.~\ref{eq:Yk_transport_u} to provide a temporal history of a fluid element evolving through a premixed flame subjected to an unsteady mixing rate.

The model is implemented in the PDRs solver~\cite{mueller2020physically,pdrs}. Chemical kinetics are modeled using the mechanism from Gotama et al.~\cite{gotama2022measurement} (31 species and 165 reactions) for AHN combustion and GRI-Mech 3.0~\cite{smith2011gri} (53 species and 325 reactions) for methane combustion. The $\Lambda$ space is discretized onto a 256 points non-uniform grid for the global flame structure. For a flame element, the progress variable step $d\LambdaAst$ is determined by solving Eq.~\ref{eq:maniL_mod2}.

\section{Characteristics of reactive nitrogen emissions}
\label{sec:pcNH3_CH4}

The partially cracking of ammonia ($\ammonia$) is obtained by following the reaction:
\begin{equation}
    2 \ammonia \rightarrow 3\hydrogen+\nitrogen,
    \label{eq:pca_reaction}
\end{equation}
\noindent
so the final fuel mixture is
\begin{equation}
    (1-\alpha)\ammonia + \frac{3\alpha}{2}\hydrogen+\frac{\alpha}{2}\nitrogen,
    \label{eq:pca_mixture}
\end{equation}
\noindent
where $\alpha$ indicates the $\ammonia$ cracking or dissociation percentage. For $\alpha=60\%$, the laminar flame speed of AHN is similar to methane \cite{jiang2020updated,wiseman2021comparison}. Simulations are carried out at gas turbine conditions, with inlet temperature $T_{\rm u}=750\rm\:K$, pressure $p=10\rm\:bar$, equivalence ratios in the range $\phi\in [0,2]$, and three AHN cracking percentages $\alpha=40\%$, 60\%, and 80\%.

\subsection{Emissions at thermodynamic equilibrium}
\label{subsec:equilibrium}

\begin{figure}[t]
\centering
\includegraphics[width=77mm]{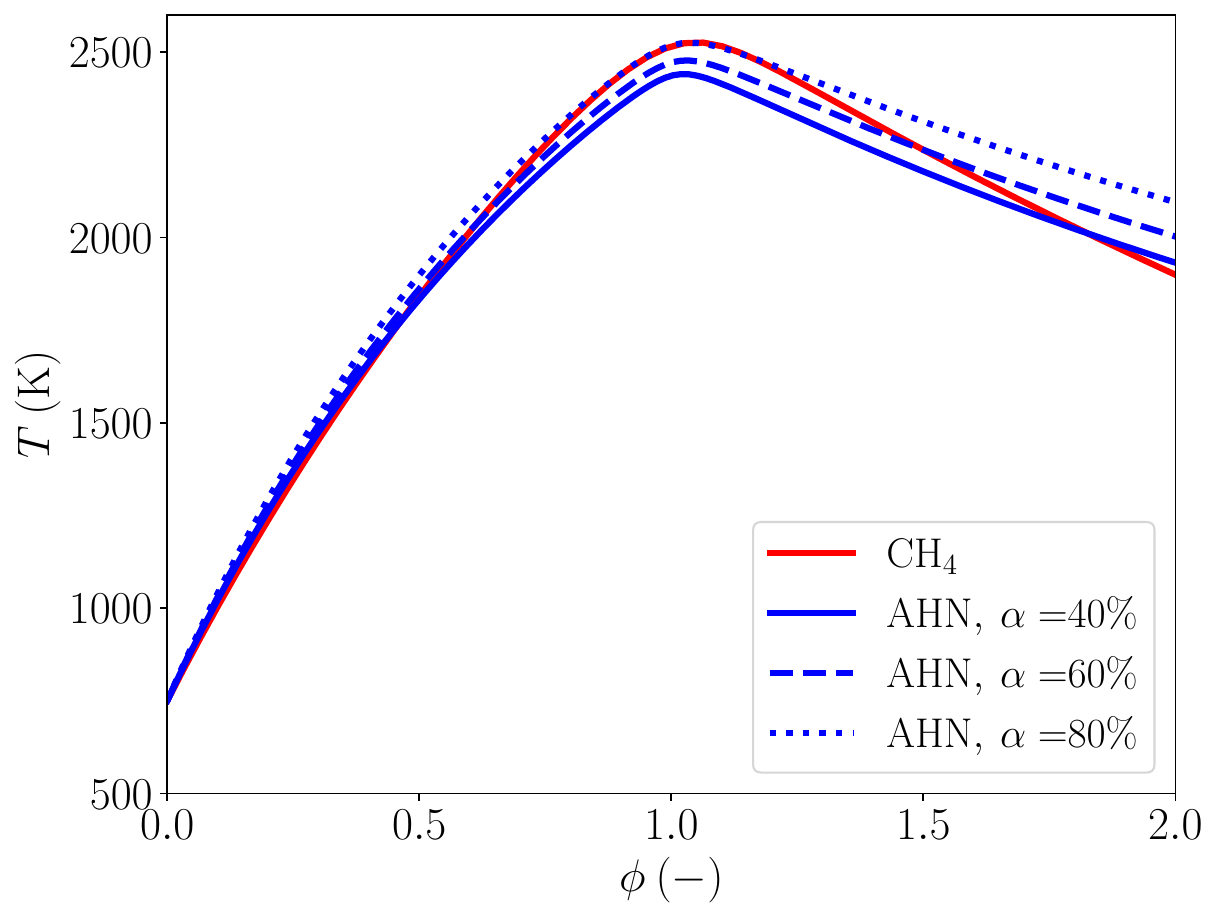}
\caption{Temperature at thermodynamic equilibrium as a function of the equivalence ratio $\phi$. Blue lines: AHN with different cracking percentages $\alpha$. Red line: methane ($\methane$).}
\label{fig:T_vs_phi}
\end{figure}

Thermodynamic equilibrium calculations of AHN- and $\methane$-air combustion are first examined.  Equilibrium temperatures for different inlet mixture composition are shown in Fig.~\ref{fig:T_vs_phi}. AHN temperatures (blue lines) are similar to those of methane (red line) at very fuel-lean mixtures. Temperature differences are observed closer to stoichiometric mixture and for fuel-rich mixtures since the heats of combustion are different for the various fuels.

\begin{figure}[t]
\centering
\includegraphics[width=77mm]{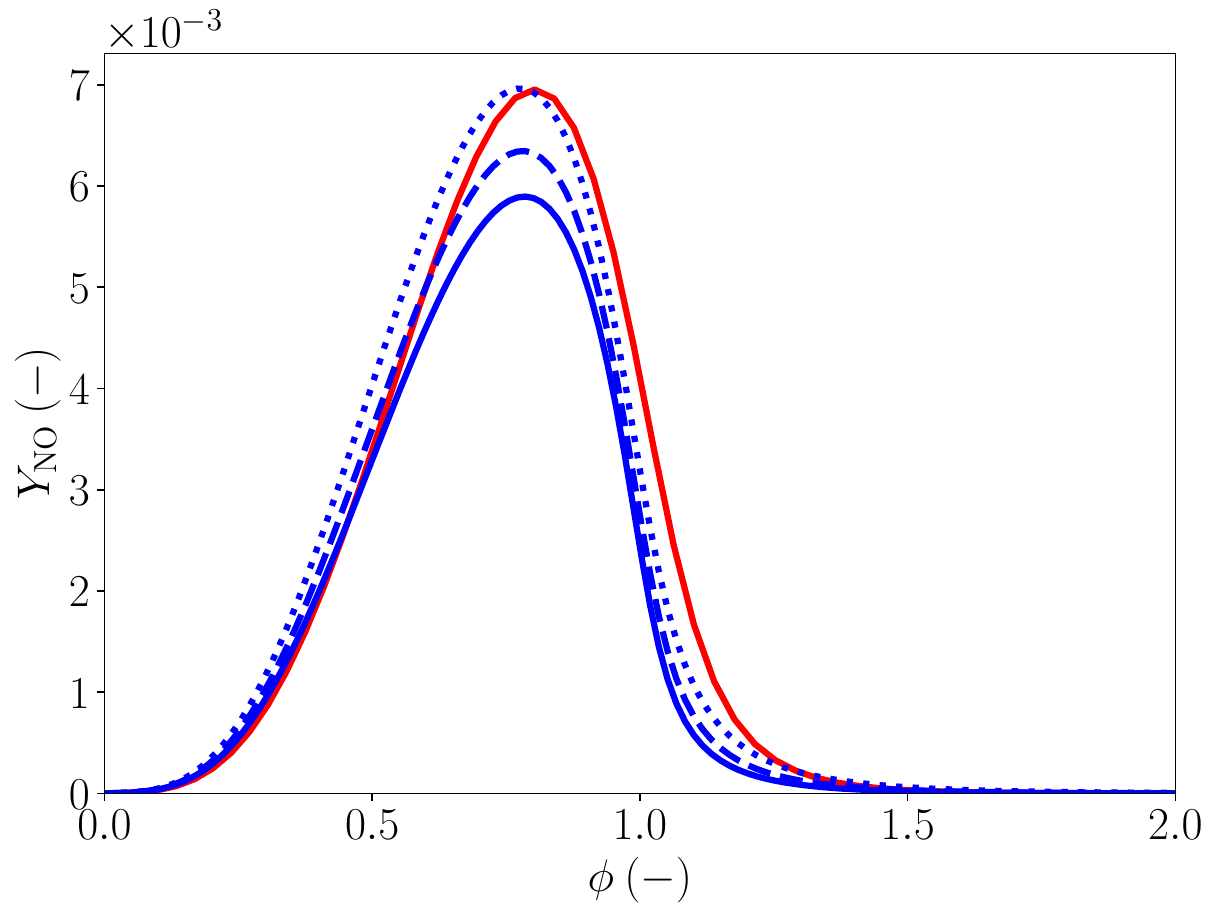}
\caption{$\NO$ mass fraction at thermodynamic equilibrium as a function of the equivalence ratio $\phi$, similar to Fig.~\ref{fig:T_vs_phi}.}
\label{fig:YNO_vs_phi}
\end{figure}

\begin{figure}[t]
\centering
\includegraphics[width=77mm]{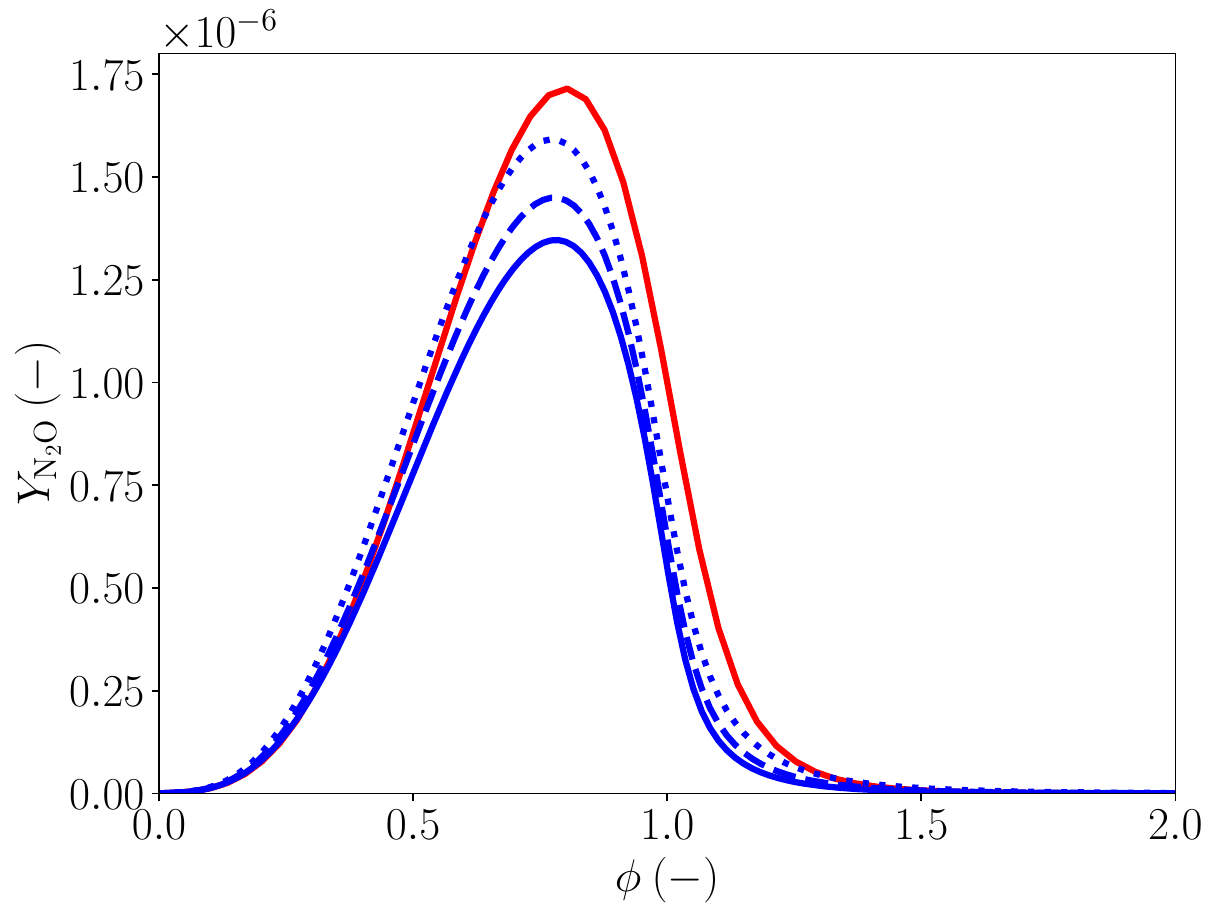}
\caption{$\NtwoO$ mass fraction at thermodynamic equilibrium as a function of the equivalence ratio $\phi$, similar to Fig.~\ref{fig:T_vs_phi}.}
\label{fig:YN2O_vs_phi}
\end{figure}

$\NO$ mass fractions are plotted in Fig.~\ref{fig:YNO_vs_phi} for both AHN (blue lines) and methane (red line) at thermodynamic equilibrium. $Y_{\NO}$ emissions are maximized at $\phi\approx0.75$, and AHN shows comparable $\NO$ emissions to methane, more or less consistent with the differences in temperature in Fig.~\ref{fig:T_vs_phi}. Similar trends and conclusions can be observed for $\NtwoO$, shown in Fig.~\ref{fig:YN2O_vs_phi}, although, counter-intuitively, methane actually has higher emissions than AHN.

These results are seemingly contrary to practice. For methane, reactive nitrogen emissions are lowest in practice for overall fuel-lean systems. For AHN, the equilibrium emissions are not any worse than hydrocarbons. The simple fact of the matter is that thermodynamics alone cannot be used to explain reactive nitrogen emissions in practical systems, for thermodynamic equilibrium is never reached with finite residence time. Since equilibrium considerations alone cannot explain emissions, the approach to equilibrium must be fundamentally different between AHN combustion and methane combustion that leads to their different emissions characteristics (higher than equilibrium for AHN and lower than equilibrium for methane) and practical combustion strategies. While this discussion is somewhat obvious, emphasizing the fundamentally different kinetic features of these fuels is important to motivate fundamentally different combustion processes.

\subsection{Emissions at finite residence time}
\label{subsec:emissions_frt_sf}

Manifold solutions for AHN- and $\methane$-air unstretched flames at gas turbine conditions for different equivalence ratios above the fuel-lean extinction limit $\phi\in[0.4,2]$ are computed. Multiple cracking percentages are considered for the AHN-air flames as before, $\alpha=40\%$, 60\%, and 80\%. Then, a reactive particle $\LambdaAst$ is tracked starting at $\Lambda=0$ for a finite residence time of $10\rm\:ms$, typical of gas turbine residence times.

\begin{figure}[t]
\centering
\includegraphics[width=77mm]{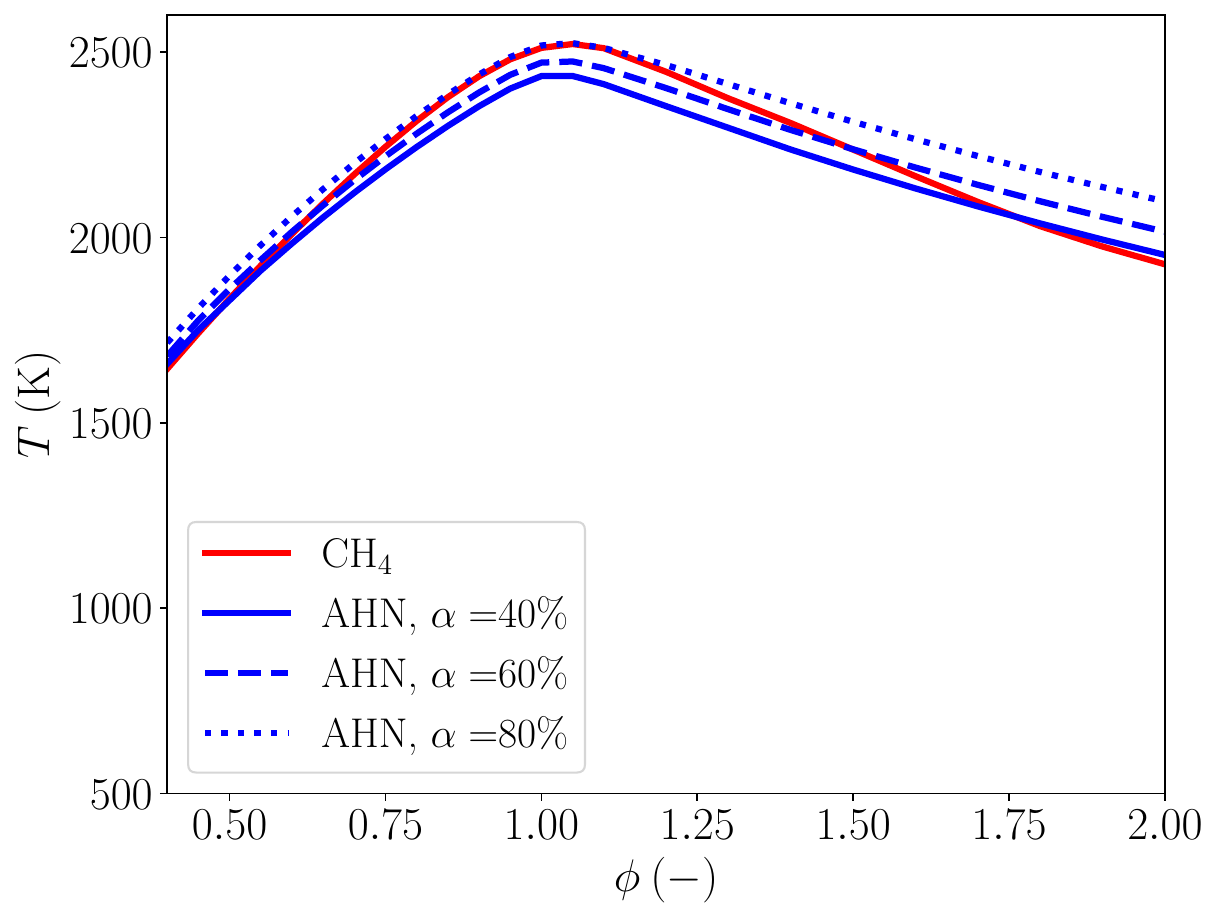}
\caption{Temperature at a finite residence time $t=10\rm\:ms$ as a function of the equivalence ratio $\phi$, at gas turbine conditions and for different cracking percentages $\alpha$. Line styles are the same as Fig.\ref{fig:T_vs_phi}.}
\label{fig:T_multi_a_multi_h2p_vs_phi}
\end{figure}

\begin{figure}[t]
\centering
\includegraphics[width=77mm]{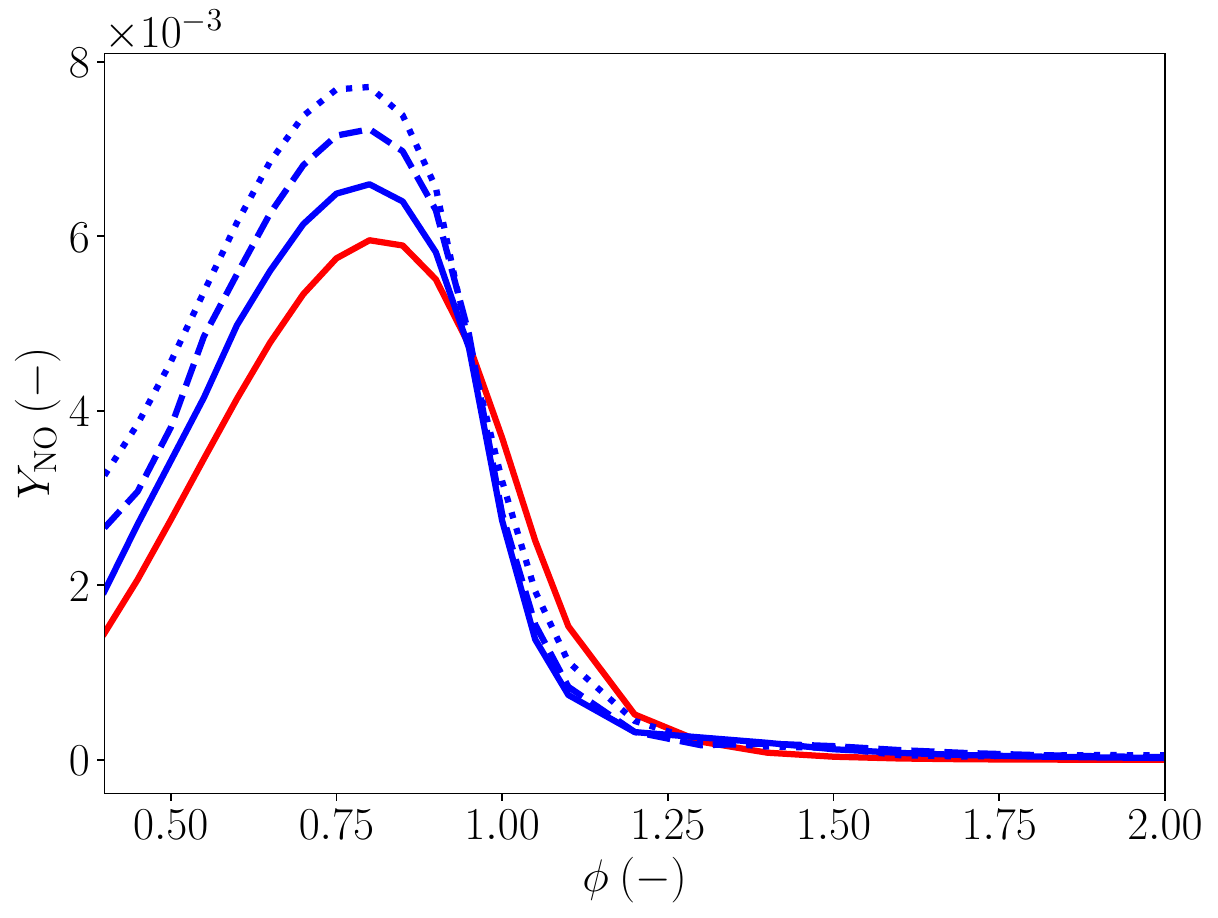}
\caption{$\NO$ mass fraction at a finite residence time $t=10\rm\:ms$ as a function of the equivalence ratio $\phi$, similar to Fig.~\ref{fig:T_multi_a_multi_h2p_vs_phi}.}
\label{fig:YNO_multi_a_multi_h2p_vs_phi}
\end{figure}

Figures~\ref{fig:T_multi_a_multi_h2p_vs_phi} and ~\ref{fig:YNO_multi_a_multi_h2p_vs_phi} show the final temperature and $\NO$ mass fraction versus equivalence ratio at $10\rm\:ms$. Results show the same trends as observed in Fig.~\ref{fig:T_vs_phi}, which explains that such flame features are associated to relatively short time scales. However, $\NO$ mass fraction differs to what is observed in Fig.~\ref{fig:YNO_vs_phi}, as its chemical reaction progress is slower than flame time scales. $\NO$ production in $\methane$-air flame is lower than in the AHN-air flames, especially for fuel-lean conditions ($\phi < 0.9$). For fuel-rich flames, AHN-air flames are comparable to $\methane$-air flames. More telling, $\NO$ emissions in the $\methane$-air flame are lower than equilibrium, while, for the AHN-air flames, they are higher than equilibrium. In general, for AHN-air flames, longer residence times will lead to a decrease in $\NO$ mass fraction, as they approach to equilibrium, and for $\methane$-air flames the opposite. Similar behavior happens with $\NtwoO$ emissions (not shown), which is negligible in $\methane$-air flames with respect to AHN flames.

\begin{figure}[t]
\centering
\includegraphics[width=77mm]{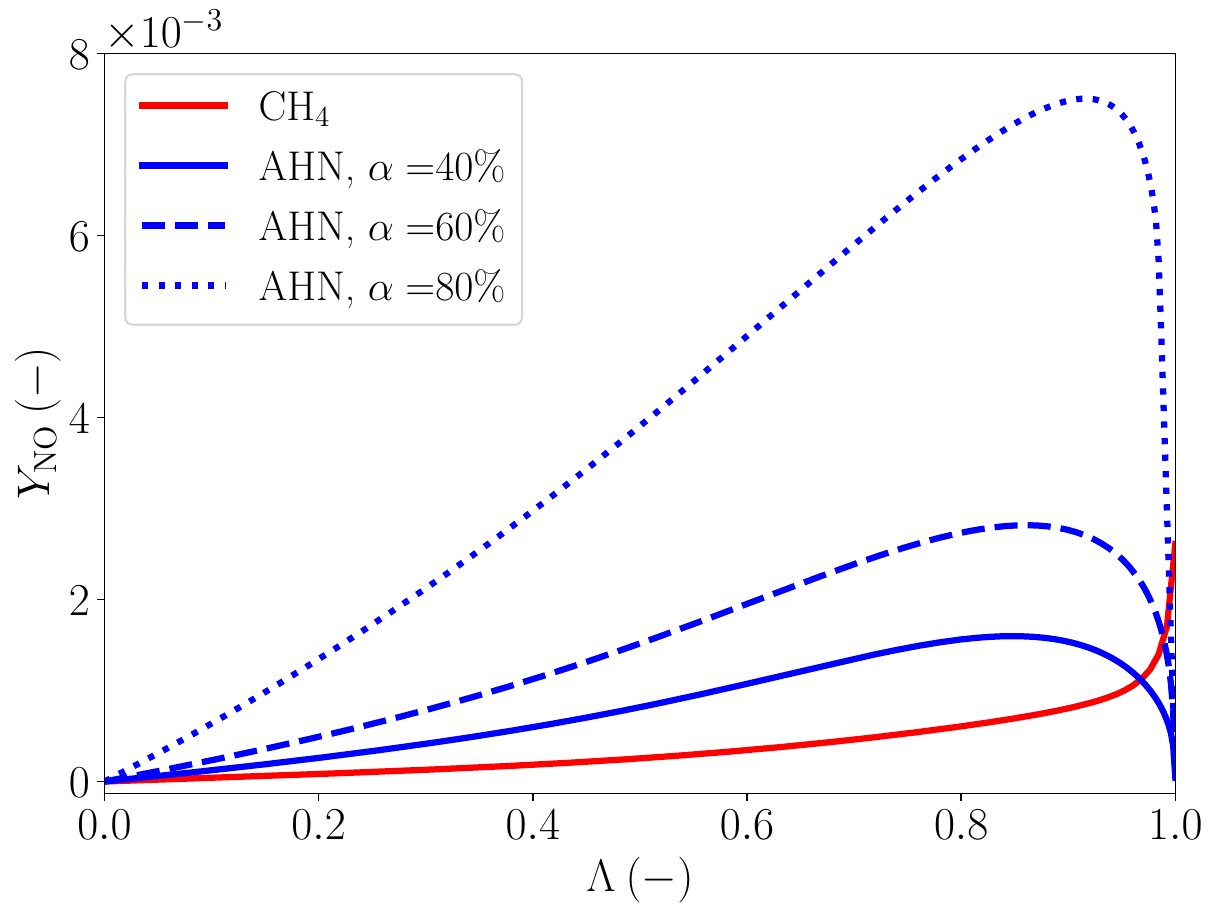}
\caption{$\NO$ mass fraction profiles in progress variables space for unstretched premixed flames at gas turbine conditions, with $\phi=0.45$ for $\methane$ (red) and $\phi=1.4$ for AHN (blue) with three different cracking percentages $\alpha$.}
\label{fig:nh3_L_YNO_multi_h2p_LgL}
\end{figure}

To understand the approach to equilibrium and the behavior of reactive nitrogen emissions at finite residence times, the (equilibrium) manifold solutions for AHN- and $\methane$-air unstretched flames are further analyzed. Typical gas turbine inlet mixtures are chosen: $\phi=0.45$ for $\methane$-air and $\phi=1.4$ (first stage) for AHN-air at three cracking extents, where $T_{\rm u}$ and $p$ are kept the same as before. The $\NO$ mass fraction profile is plotted in Fig.~\ref{fig:nh3_L_YNO_multi_h2p_LgL}. The $Y_{\NO}$ profile in the $\methane$-air flame shows low emissions up to $\Lambda\approx0.9$ and abruptly increases as the $\Ntwo$ bond breaks in the hot burned gases. In the AHN-air combustion, $Y_{\NO}$ forms much earlier in the flames, peaking within the intense reaction zone and then ``relaxing'' toward equilibrium ($0.9\leq\Lambda\leq1$). The maximum NO mass fraction is a strong function of the cracking percentage, increasing with the cracking percentage for the values shown here (at higher cracking yet, the peak reactive nitrogen mass fractions will decrease with cracking percentage as the fuel mixture approaches a mixture of only hydrogen and nitrogen).

The approach to equilibrium then dictates the practical combustion strategies. For methane-air combustion, not too long residence times are desirable since reactive nitrogen emissions form only very close to equilibrium. On the other hand, for AHN-air combustion, long residence times are desirable to get as close to thermodynamic equilibrium as possible and minimize ``unrelaxed'' emissions. However, longer residence times incur additional capital costs and would also increase heat losses so would not necessarily be desirable. Therefore, a combustion process that could accelerate the relaxation to equilibrium without increasing residence time would be highly desirable for minimizing reactive nitrogen emissions in AHN combustion.

\begin{figure*}[t]
\centering
\includegraphics[width=\textwidth]{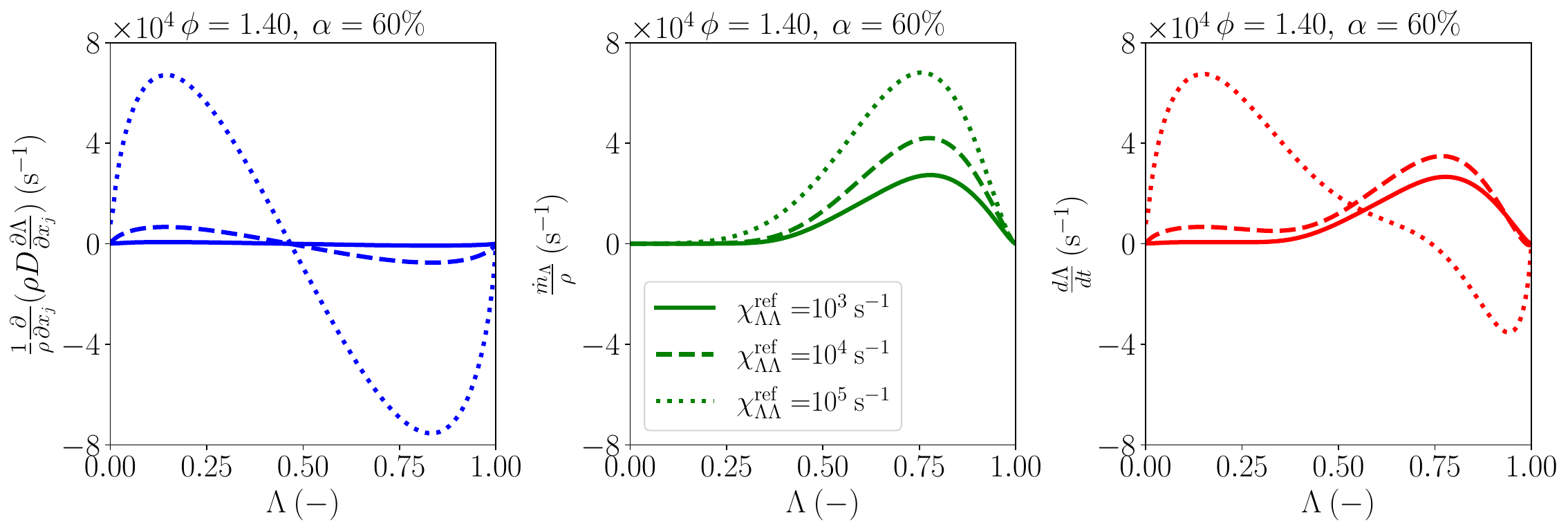}
\caption{Right-hand-side of the progress variable transport equation (Eq.~\ref{eq:maniL_mod}) in progress variable space: mixing term (left), chemical term (center), and total (right). Three reference dissipation rates $\chiLLref$ are considered: $10^{3}\rm\:s^{-1}$ (solid), $10^{4}\rm\:s^{-1}$ (dashed), and $10^{5}\rm\:s^{-1}$ (dotted).}
\label{fig:pcnh3_3figs_dLdt=0}
\end{figure*}
\begin{figure}[h]
\centering
\includegraphics[width=77mm]{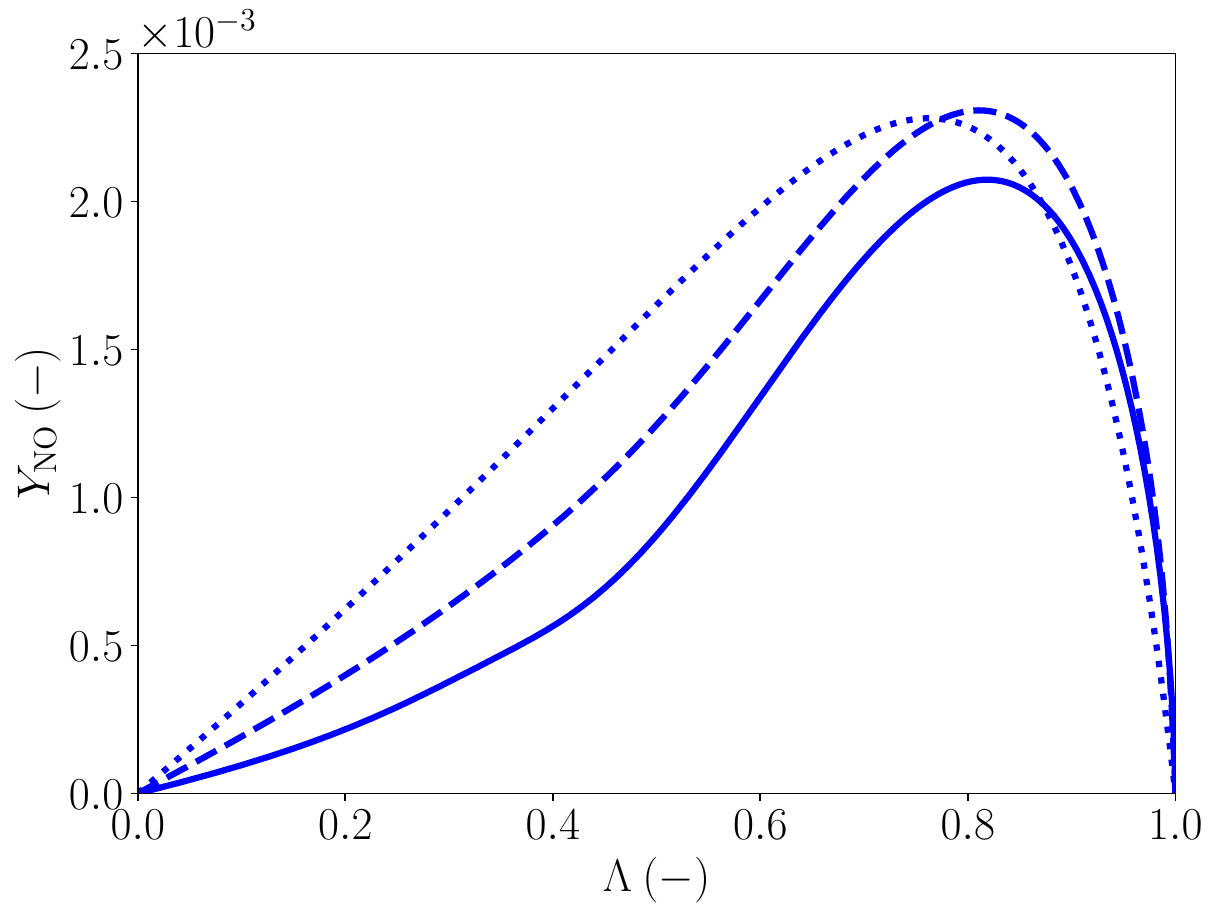}
\caption{$\NO$ mass fraction profiles in progress variables space for AHN premixed stretched flames at gas turbine conditions and three different reference dissipation rates $\chiLLref$, same as Fig.~\ref{fig:pcnh3_3figs_dLdt=0}.}
\label{fig:pcnh3_L_YNO}
\end{figure}

\section{Rush-to-equilibrium concept}
\label{sec:rush2eq}

As discussed in the previous section, accelerating the approach to equilibrium is critical to minimizing reactive nitrogen emissions in AHN combustion. To develop such a combustion strategy, the fundamental balance in premixed combustion must be analyzed. Terms on the right-hand side of the progress variable transport equation of Eq.~\ref{eq:maniL_mod} govern the increase in the progress variable of a reacting flow particle and are shown in Fig.~\ref{fig:pcnh3_3figs_dLdt=0} for AHN combustion at the same conditions considered previously for different values of the mixing rate $\chiLLref$ and indicate the forcing on the progress variable. For the mixing term (left), the term is positive at low progress variable and negative at high progress variable. The positive values at low progress variable are responsible for transport from the reaction zone through the preheat zone. Since chemistry is not present at this stage, passive transport through the preheat zone occurs. Increasing the reference dissipation rate significantly increases mixing effects in the flame. The chemical term (center) is exclusively positive and largest at higher progress variables in the reaction zone. The balance of this competition is shown in the total right-hand-side (right). For all mixing rates, the (negative) mixing at high progress variable inhibits the chemistry's progress, making the net progress rate is lower than chemistry, making the approach to equilibrium ``slower''. In fact, at the highest mixing rates, the total term can become negative, resulting in a ``frozen'' state at the zero crossing of the total ($\Lambda\approx0.74$ for the highest dissipation rate in the plots), far from equilibrium, and with very high unrelaxed $\NOx$ emissions (although such a flame would never be practically realized for obvious reasons). These $\NOx$ emissions are shown in Fig.~\ref{fig:pcnh3_L_YNO} for the three cases. Indeed, as the residence time is increased, the equilibrium conditions will be reached except for the fast mixing case (dotted line), where a mass fraction higher than $2\times10^{-3}$ is reached at the ``frozen'' state.

This fundamental balance then motivates a novel rush-to-equilibrium combustion concept to get as close to equilibrium as fast as possible. As a reacting flow particle moves through the flame, at low progress variable, fast mixing rates are desirable, promoting the increase in progress variable, but, at high progress variable, slow mixing rates are desirable, avoiding suppression of chemistry's tendency to increase the progress variable. Therefore, as a reacting flow particle travels through the flame, a temporally decaying (in a Lagrangian sense) mixing rate would be desirable. For proof-of-concept, the following temporally evolving mixing rate is considered:
\begin{equation}
     \chiLLref\left(t\right) = \frac{\chiLLnot}{1 + a\:\chiLLref t},
     \label{eq:chiLLref_t}
\end{equation}
where an initial reference dissipation rate $\chiLLref(0)=\chiLLnot$ and a decay rate $a$ are prescribed. This decaying mixing rate will promote a faster approach to equilibrium, so a fixed residence time of $10\rm\:ms$, typical of gas turbine combustors, will be considered.

\section{Evaluation of the rush-to-equilibrium concept}
\label{sec:eval_concept}

In Section~\ref{sec:pcNH3_CH4}, it was observed that reactive nitrogen emissions are higher at high cracking percentages, because the reacting mixtures have higher final temperatures (see Figs.~\ref{fig:T_vs_phi} and \ref{fig:T_multi_a_multi_h2p_vs_phi}) due to $\hydrogen$ reactivity. To provide a fairer comparison with cracking percentages, the evaluation of the rush-to-equilibrium concept is made by considering the equilibrium temperature (roughly) the same for all mixtures, this is, by keeping enthalpy the same for all mixtures. Therefore, the initial temperature for the mixtures are chosen such that the enthalpy of the mixture is equal to that of the case with $\alpha=40\%$, $p=10\rm\:bar$, and $\phi=1.4$.

\begin{figure*}[t]
\centering
\includegraphics[width=\textwidth]{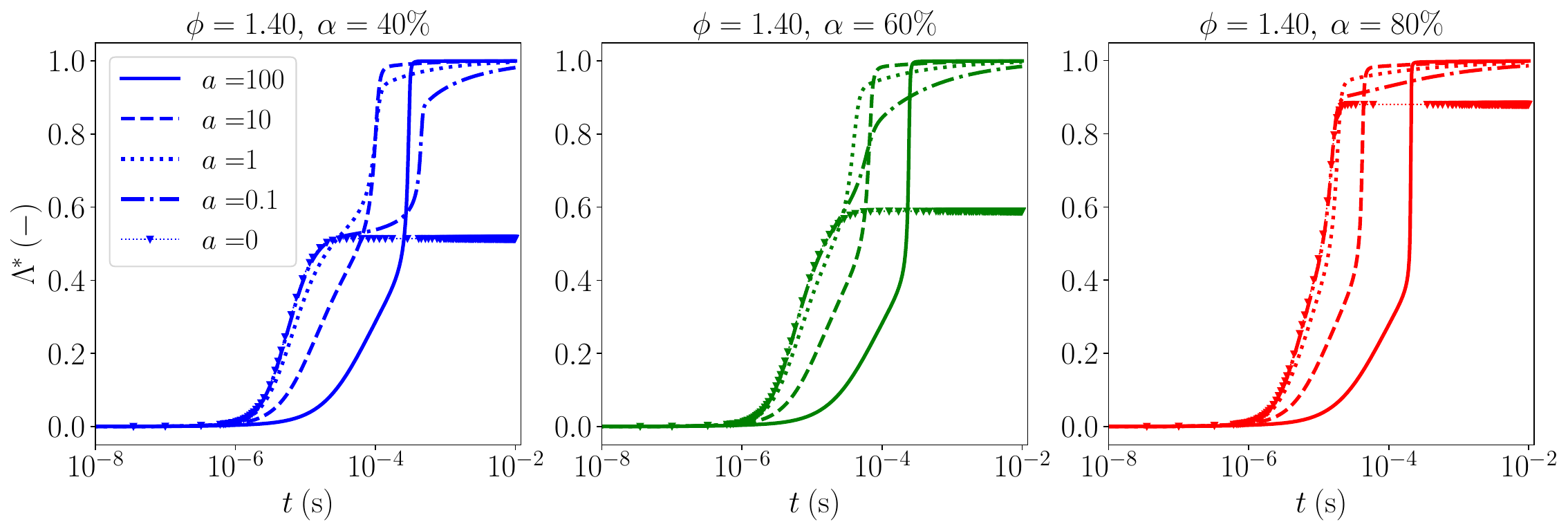}
\caption{Effect of decaying mixing rate on the temporal evolution of the progress variable of a reacting flow particle $\LambdaAst$ in AHN-air premixed flames with $\phi=1.4$ and $\ammonia$ cracking percentages $\alpha=40\%$ (left, blue), $60\%$ (center, green), and $80\%$ (right, red), considering initial reference dissipation rate $\chiLLnot=10^5\rm\:s^{-1}$ and multiple decay rates $a$.}
\label{fig:LambdaAst_vs_t}
\end{figure*}

Figure~\ref{fig:LambdaAst_vs_t} shows the results of the temporal evolution of the progress variable of the reacting flow particle $\LambdaAst$ for initial mixing rate $\chiLLnot=10^{5}\rm\:s^{-1}$, multiple cracking extents, and different decay rates. As expected, the decaying mixing rate allows the fluid element to move from the initial unburned mixture to the reacting stage to (near) equilibrium conditions. For this relatively fast initial mixing rate, a constant dissipation rate ($a=0$) leads to a ``frozen'' condition, as discussed above. Note that the chemical source terms increase with cracking extent, so the ``frozen'' condition is closer to equilibrium with increasing cracking extent. As the decay rate increases, for all cracking extents, the time needed to reach near equilibrium condition decreases, and the thermochemical state is nearly to thermodynamic equilibrium with the same fixed residence time. The transit time through the preheat zone is longer with the fastest decay rates. Although not shown here, for slower initial mixing rates, if the decay rate is too fast, transport will not be strong enough to provide the initial increase in the progress variable, and the flow particle will be stuck at a very low progress variable state since the homogeneous ignition delay time for this mixture is very long.

\begin{figure*}[!t]
\centering
\begin{minipage}{\textwidth}
  \centering
  \includegraphics[trim={0mm 11mm 0mm 0mm},clip,width=0.93\textwidth]{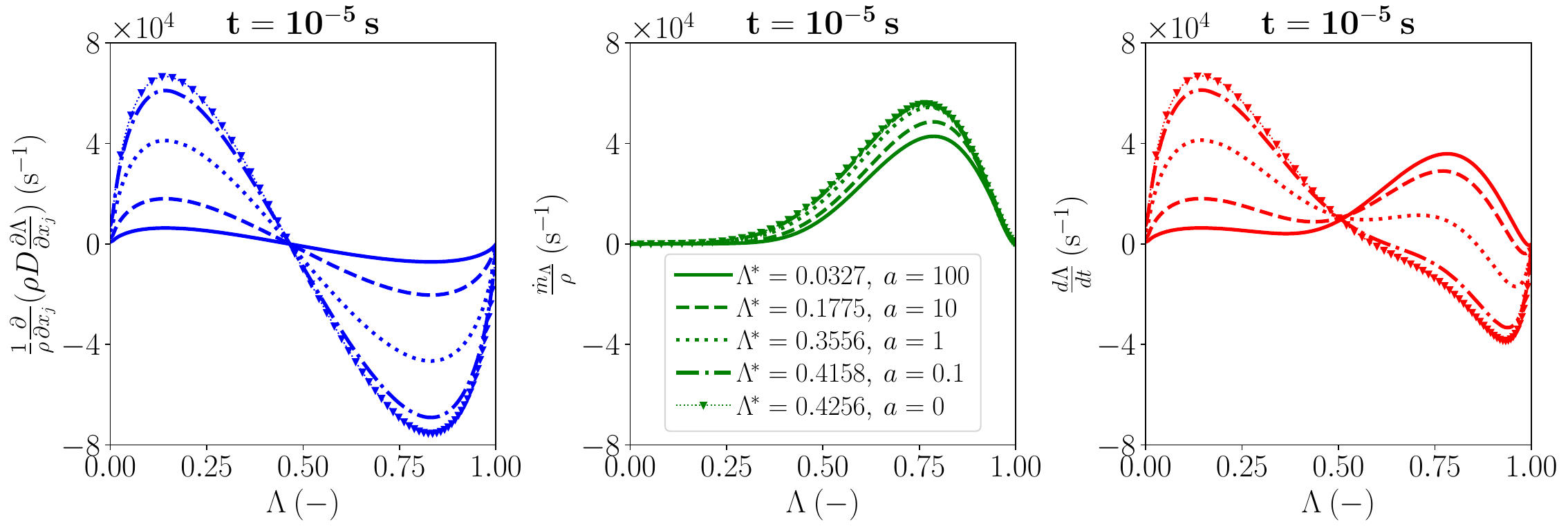}
\end{minipage}
\begin{minipage}{\textwidth}
  \centering
  \includegraphics[trim={0mm 11mm 0mm 0mm},clip,width=0.93\textwidth]{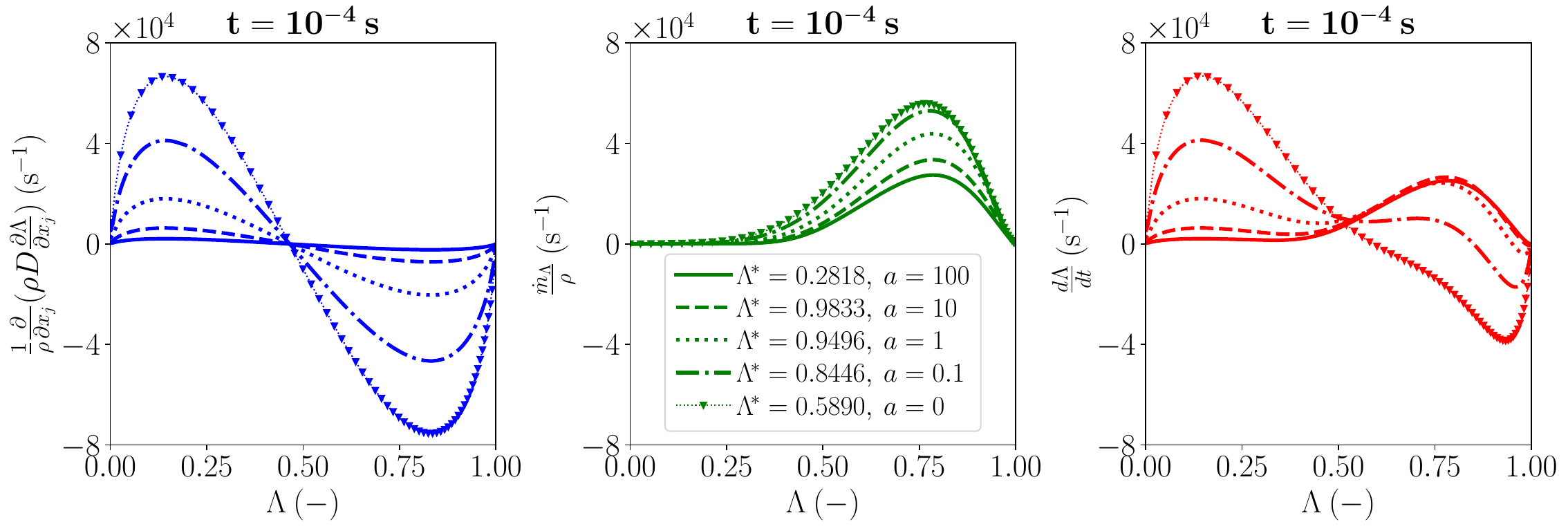}
\end{minipage}
\begin{minipage}{\textwidth}
  \centering
  \includegraphics[trim={0mm 11mm 0mm 0mm},clip,width=0.93\textwidth]{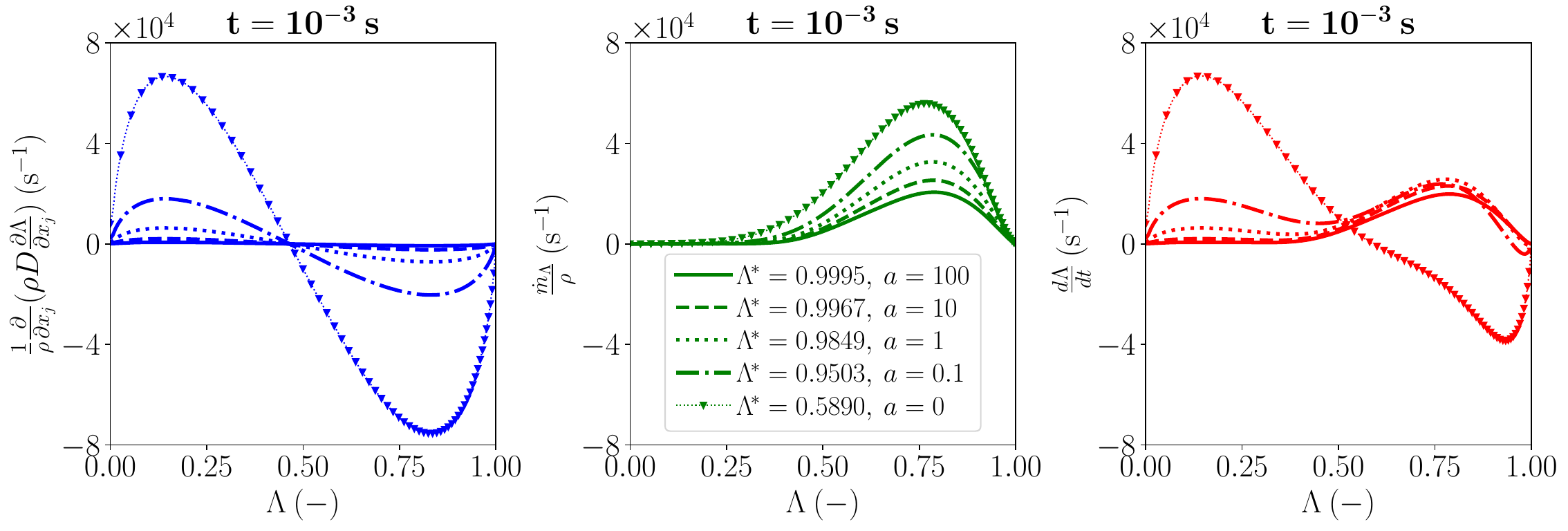}
\end{minipage}
\begin{minipage}{\textwidth}
  \centering
  \includegraphics[trim={0mm 0mm 0mm 0mm},clip,width=0.93\textwidth]{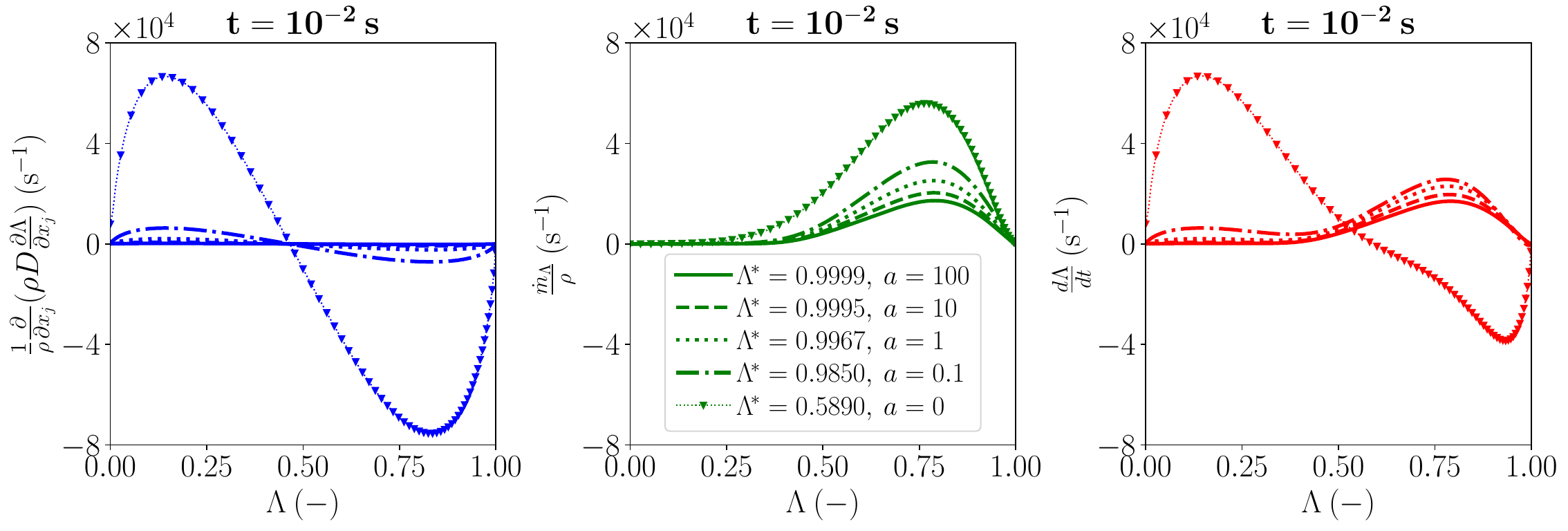}
\end{minipage}
\caption{Effect of decaying mixing rate on progress rate terms (Eq.~\ref{eq:maniL_mod}) in progress variable space for the global flame structure ($\phi=1.4$, $\alpha=60\%$, $\chiLLnot=10^{5}\rm\:s^{-1}$) at four different snapshots: mixing (left), chemical (middle), and total (right) terms. $\LambdaAst$ are from Fig.~\ref{fig:LambdaAst_vs_t}.}
\label{fig:pcnh3_3figs_dLdt_t} 
\end{figure*}

A deeper insight on how the decay rate affects the global flame structure is shown in Fig.~\ref{fig:pcnh3_3figs_dLdt_t}, by analyzing the temporal evolution of the progress variable terms from Eq.~\ref{eq:maniL_mod} for the flame with $\phi=1.4$ and cracking percentage $\alpha=60\%$, where $a$ varies from 0 (no decay) to 100 (unrealistic excessive decay). Similar conclusions can be obtained with other cracking percentages. Four instants $t=10^{-5}\rm\:s$, $10^{-4}\rm\:s$, $10^{-3}\rm\:s$, and $10^{-2}\rm\:s$ (rows) are shown, and the position of the reactive element $\LambdaAst$ from Fig.~\ref{fig:LambdaAst_vs_t} is indicated for each case. The initial state ($t=0$) for all cases corresponds to $\chiLLnot=10^{5}\rm\:s^{-1}$ and coincides with the case with no decay $a=0$ (line with symbols), which conserves the same shape in all four rows as expected. For $t=10^{-5}$, mixing (left column) significantly decreases for fast decays whereas chemistry (middle column) is barely affected, giving positive balance in the total progress (right column) for rather fast decays compared to slow to no decay cases. A fast decay affect the initial progress significantly: for a very fast decay ($a=100$), the progress of $\LambdaAst$ is about 11 times smaller than the moderate decay ($a=1$) and about 13 times smaller with respect to no decay ($a=0$), meaning initial stage for fast decays versus almost midway stages for slower decays. Nevertheless, the case with no decay reached the ``frozen'' state due to zero balance between mixing and chemistry as can be observed in the following instants. At $t=10^{-4}\rm\:s$, the decaying mixing affects chemistry at a smaller proportion compared to mixing rate for moderate to fast decay, where only a half reduction of the chemistry term is observed. The total progress becomes more positive for decaying mixing. The reactive element is still located in the first half of the progress variable domain for very fast decay compared to other cases at about 3.3 times smaller with respect to moderate decay ($a=1$) and 2.5 times smaller compared to no decay ($a=0$). This is followed by the abrupt progress at fast decay, where at $t=1\rm\:ms$, the reactive element in decaying cases is at 95\% reaction progress or more. Total progress profiles are positive for decaying cases, with slow mixing at the final stage and almost unaffected chemistry effects with respect to the previous state. A slight decrease of the combined effects are observed at $t=10\rm\:ms$, where the reactive particle is at near equilibrium conditions for all decaying mixing scenarios.
 
\begin{figure*}[!t]
\centering
\begin{minipage}{\textwidth}
  \centering
  \includegraphics[trim={0mm 11mm 0mm 0mm},clip,width=\textwidth]{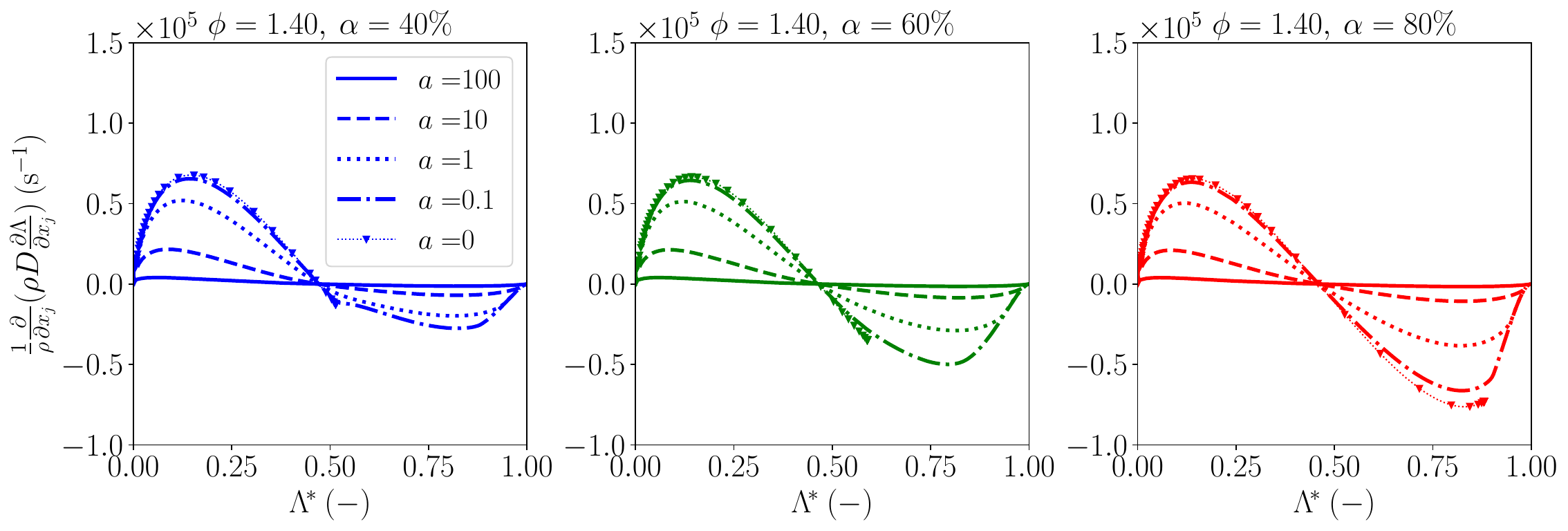}
\end{minipage}  
\begin{minipage}{\textwidth}
  \centering
  \includegraphics[trim={0mm 11mm 0mm 0mm},clip,width=\textwidth]{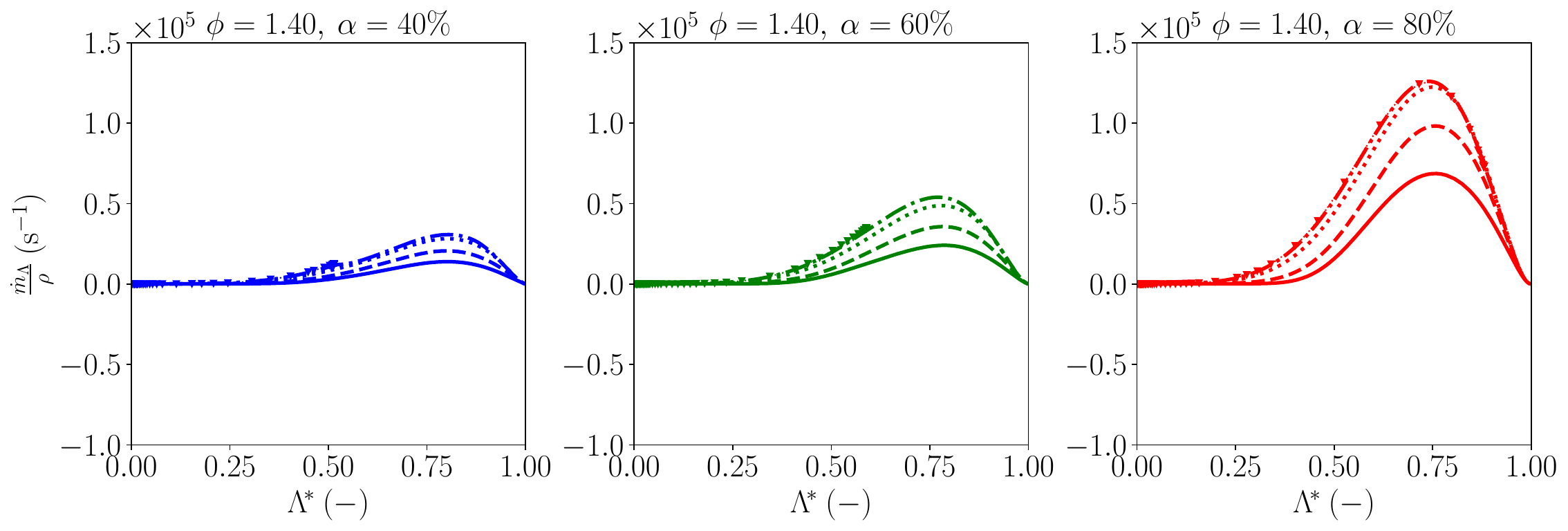}
\end{minipage}
\begin{minipage}{\textwidth}
  \centering
  \includegraphics[trim={0mm 0mm 0mm 0mm},clip,width=\textwidth]{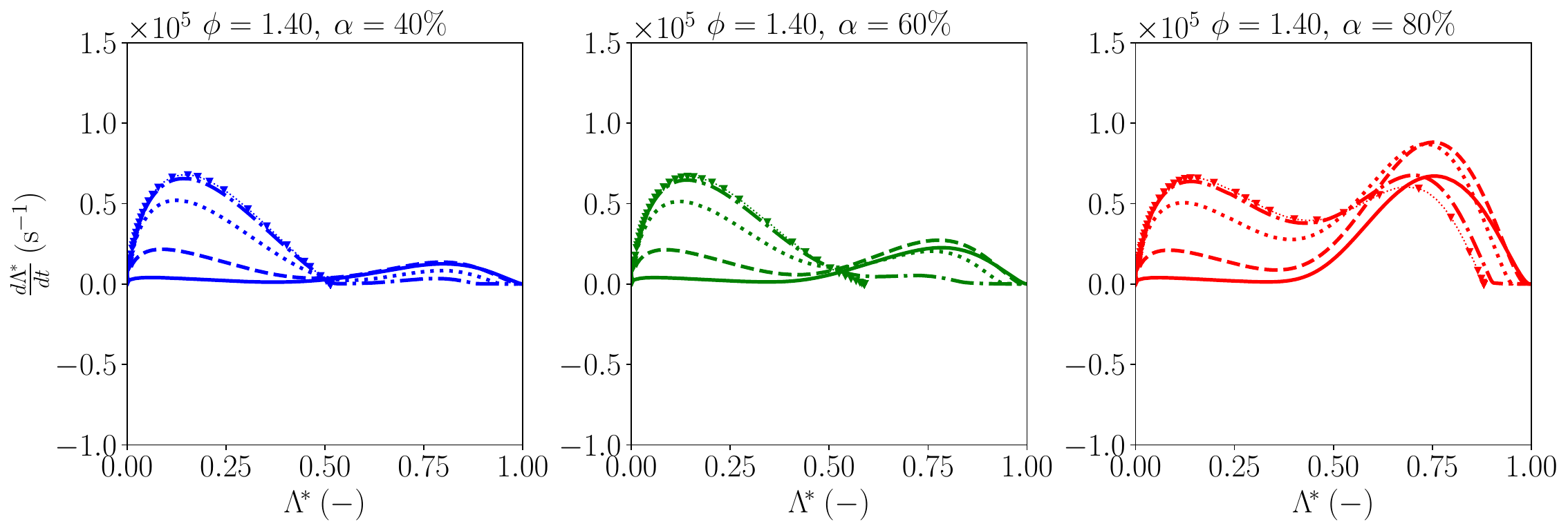}
\end{minipage}
\caption{Effect of decaying mixing rate on the evolution of progress rate terms from Eq.~\ref{eq:maniL_mod} in progress variable space for a reactive element $\LambdaAst$ in AHN flames with $\phi=1.4$ and three cracking percentages: mixing (top), chemical (middle), and total (bottom) terms. Line styles correspond to the same cases from Fig.~\ref{fig:LambdaAst_vs_t}.}
\label{fig:dLdt_rhs_chiLLi_multi_a_multi_h2p} 
\end{figure*}

To quantify the effects of cracking extent, the reactive element trajectory is analyzed in progress variable space from a Lagrangian point of view. Figure \ref{fig:dLdt_rhs_chiLLi_multi_a_multi_h2p} shows the right-hand-side terms from Eq.~\ref{eq:maniL_mod} (top and middle rows) and the total reaction progress rate $d \LambdaAst / dt$ (bottom) versus the position of the reactive element $\LambdaAst$ using the rush-to-equilibrium concept and evaluated for different decay rates $a$ and cracking percentages $\alpha$ and fixed initial mixing rate $\chiLLnot=10^{5}\rm\:s^{-1}$. The decaying mixing rate effectively allows for a positive progress rate from the beginning to the end of the reacting element trajectory (bottom row), which decomposes into a high positive mixing term in the first half (top row) and a chemically-driven second half (middle row) with low negative mixing (top row). Decay rates have significant effects in the first half of the mixing rate profiles (top row). As observed in Fig.~\ref{fig:pcnh3_3figs_dLdt_t}, chemistry is much less affected by the decaying rate compared to mixing, with almost one half difference between very fast decay and no decay cases. The total progress rate increases in the second half as the the reactivity of the mixture increases with cracking much more compared to negative mixing. The balance between mixing and chemistry elucidates that moderate to high decaying rates are preferred to reach near equilibrium conditions faster in agreement with the observations in Fig.~\ref{fig:LambdaAst_vs_t}.

\begin{figure*}[t]
\centering
\includegraphics[width=\textwidth]{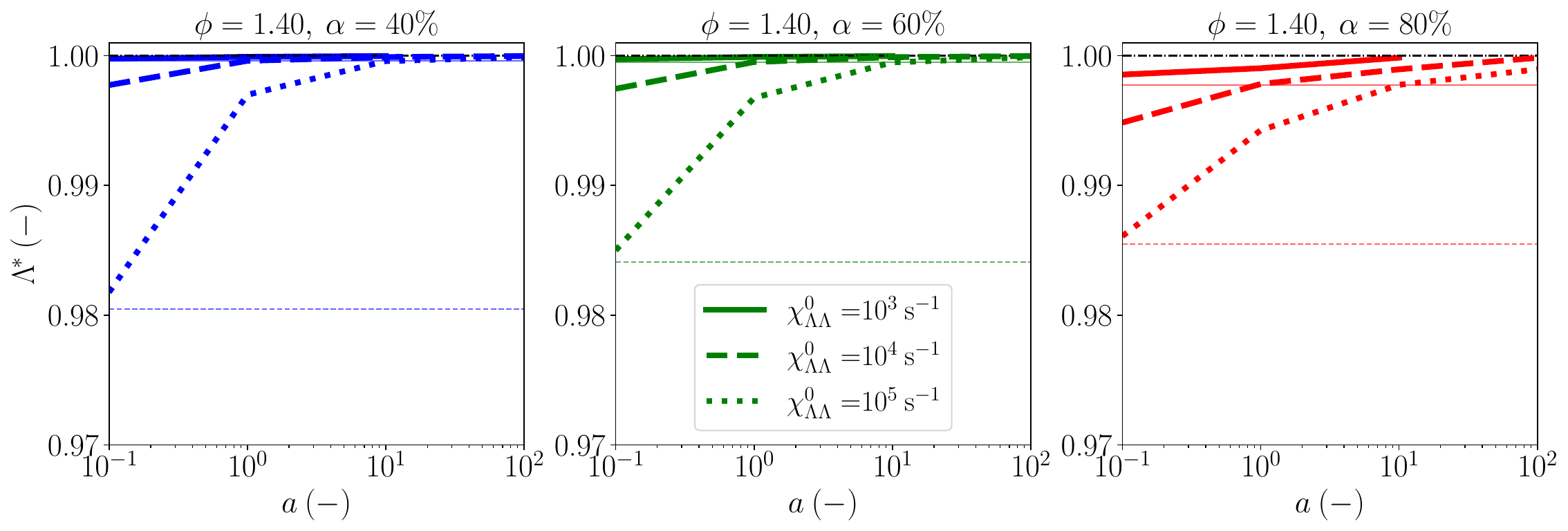}
\caption{Progress variable of the reacting flow particle $\LambdaAst$ at a 10 ms residence time as a function of the mixing rate decay rate $a$. Three initial reference dissipation rates $\chiLLnot$ are considered: $10^{3}\rm\:s^{-1}$ (solid), $10^{4}\rm\:s^{-1}$ (dashed), and $10^{5}\rm\:s^{-1}$ (dotted). The thin transparent lines represent the same quantities but using $a=0$, and the black dash-dotted line represents thermodynamic equilibrium.}
\label{fig:H_LambdaAst_vs_a_end}
\end{figure*}

The progress variables of the reacting flow element $\LambdaAst$ at the fixed 10 ms residence time as a function of the decay rate are shown in Fig.~\ref{fig:H_LambdaAst_vs_a_end} and support the discussion above. Three initial reference dissipation rates are considered, and only cases reaching close to thermodynamic equilibrium by $t=10\rm\:ms$ are included. Results with zero decay (transparent thin lines) and the thermodynamic equilibrium condition (black line) are also included. In all cases, a decaying mixing rates promotes a closer approach to equilibrium compared to the constant mixing rate. As $\chiLLnot$ increases, the fluid element benefits more from faster mixing decay rates. As the cracking percentage increases, equilibrium is approached more slowly towards the end of the simulation for slow decays, which is consistent with Figs.~\ref{fig:LambdaAst_vs_t}--\ref{fig:dLdt_rhs_chiLLi_multi_a_multi_h2p}, but still closer to equilibrium compared to constant mixing cases.

\begin{figure*}[!t]
\centering
\begin{minipage}{\textwidth}
  \centering
  \includegraphics[trim={0mm 11mm 0mm 0mm},clip,width=\textwidth]{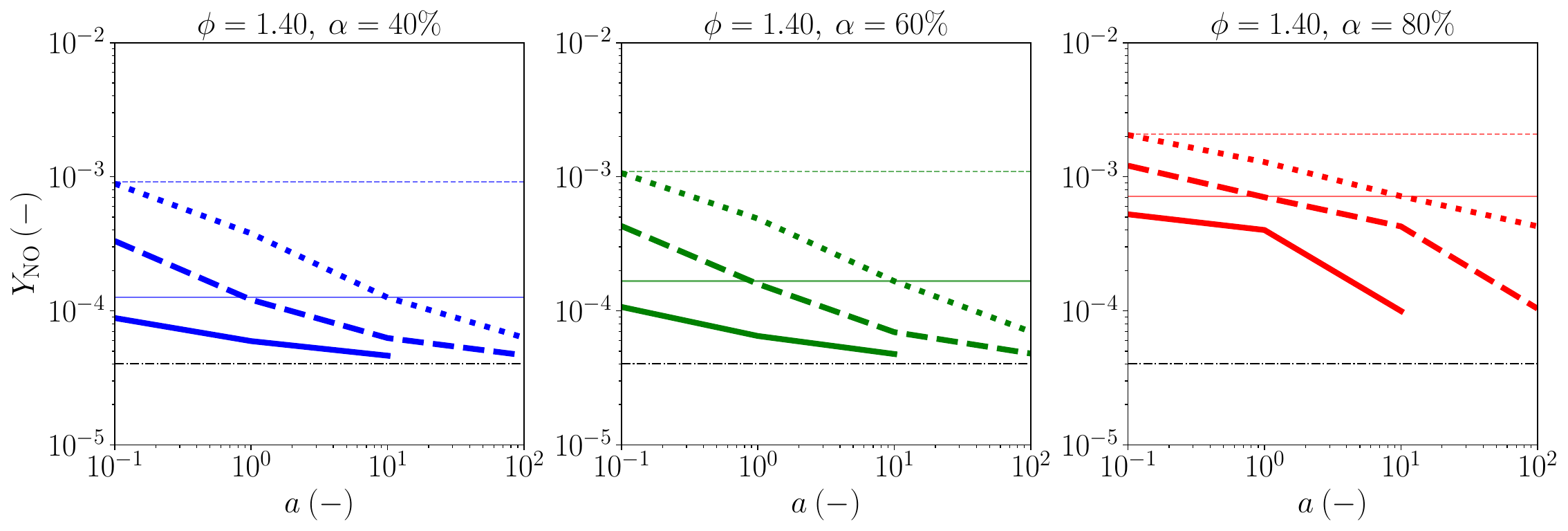}
\end{minipage}
\begin{minipage}{\textwidth}
  \centering
  \includegraphics[trim={0mm 0mm 0mm 10mm},clip,width=\textwidth]{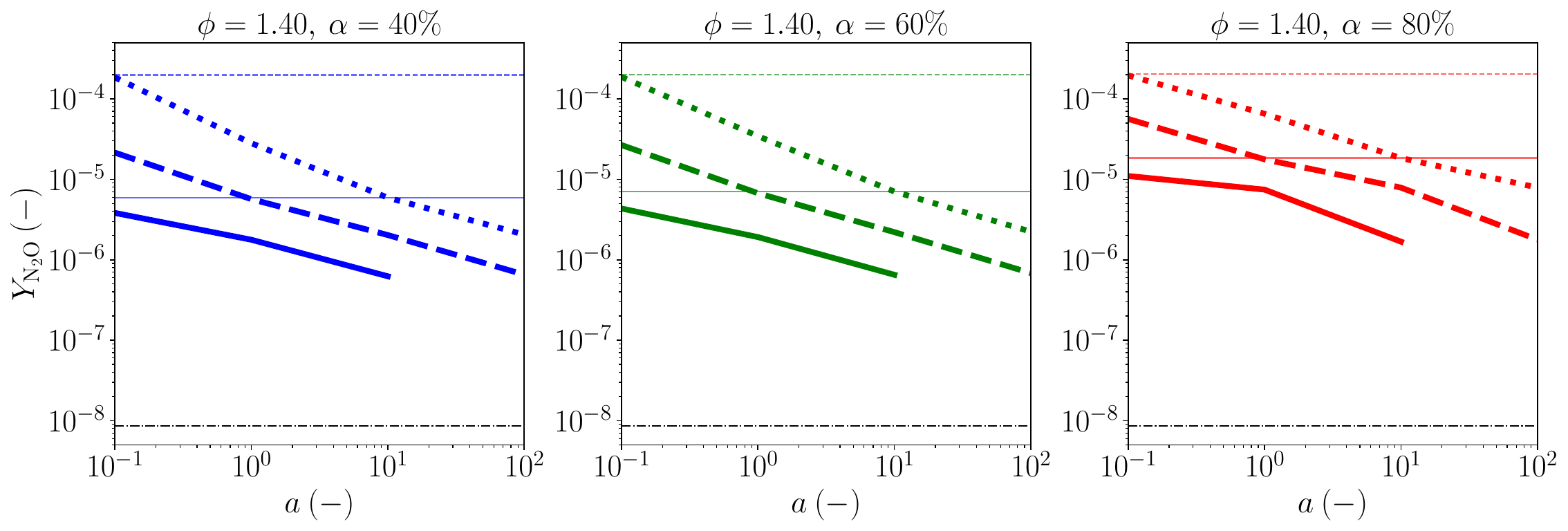}
\end{minipage}
\caption{$\NO$ (top) and $\NtwoO$ (bottom) mass fraction of the reacting flow particle at a 10 ms residence time as a function of the decay rate $a$. Line styles correspond to the same cases from Fig.~\ref{fig:H_LambdaAst_vs_a_end}.}
\label{fig:H_YNO_YN2O_vs_a_end} 
\end{figure*}

Figure~\ref{fig:H_YNO_YN2O_vs_a_end} shows the corresponding reactive nitrogen emissions of $\NO$ (top) and $\NtwoO$ (bottom) at the fixed 10 ms residence time as a function of the decay rate; note the logarithmic scale of the vertical axis. In general, $\NO$ and $\NtwoO$ emissions decrease as the reacting flow element more closely approaches thermodynamic equilibrium. Consistent with the decaying mixing rates providing a final thermochemical state closer to thermodynamic equilibrium, decaying mixing rates significantly reduce both reactive nitrogen emissions through a reduction in the unrelaxed emissions. For slower decay rates, the reactive nitrogen emissions are more sensitive to the initial mixing rate, but they are still much lower than the corresponding constant mixing rates, a factor of two or more for $\NO$ and a factor of three or more for $\NtwoO$ at the smaller cracking percentages. For instance, for moderate initial mixing rate ($\chiLLnot=10^{4}\rm\:s^{-1}$, dashed line), cracking ($\alpha=60\%$, middle column), and decay rate ($a=1$), $\NO$ emissions are about 7 times lower and $\NtwoO$ about 29 times lower compared to those with constant mixing ($a=0$). For fast initial mixing ($\chiLLnot=10^{5}\rm\:s^{-1}$, dotted lines), the emissions with constant mixing are too high as they do not reach near equilibrium conditions (lines are outside the plotting window). For the fastest decay rates, at the smaller cracking percentages, the emissions reach (nearly) their thermodynamic equilibrium values, almost entirely eliminating the unrelaxed $\NO$ emissions, but, while moderate decay rates are certainly practical (see the next subsection), these fastest decay rates may not be practical. Finally, for the same initial mixing rates and decay rates, for the highest cracking percentage, the reduction in the $\NO$ emissions is not as strong, with a factor of two or more with decreasing initial mixing rates. $\NtwoO$ is only slightly higher, with a factor of two or less with increasing initial mixing rates. This may motivate the use of \emph{less} cracked ammonia to minimize reactive nitrogen emissions. Overall, although $\NtwoO$ emissions are about two order of magnitude or more higher with respect to equilibrium conditions, these reductions are still substantial, especially for high initial mixing rates. Results considering the same initial temperature instead of the constant enthalpy are included in the Supplementary Materials. Also, the use of different decaying mixing rate profiles can significantly improve the reduction of nitrogen emissions. An example of using exponential decay is shown in the Supplementary Materials, where reactive nitrogen emissions can reach as low as equilibrium conditions. However, practical application may not be feasible, which motivates future work on the optimization of decaying mixing rate profiles to achieve low emissions.

\begin{figure*}[!t]
\centering
\begin{minipage}{\textwidth}
  \centering
  \includegraphics[trim={0mm 11mm 0mm 0mm},clip,width=\textwidth]{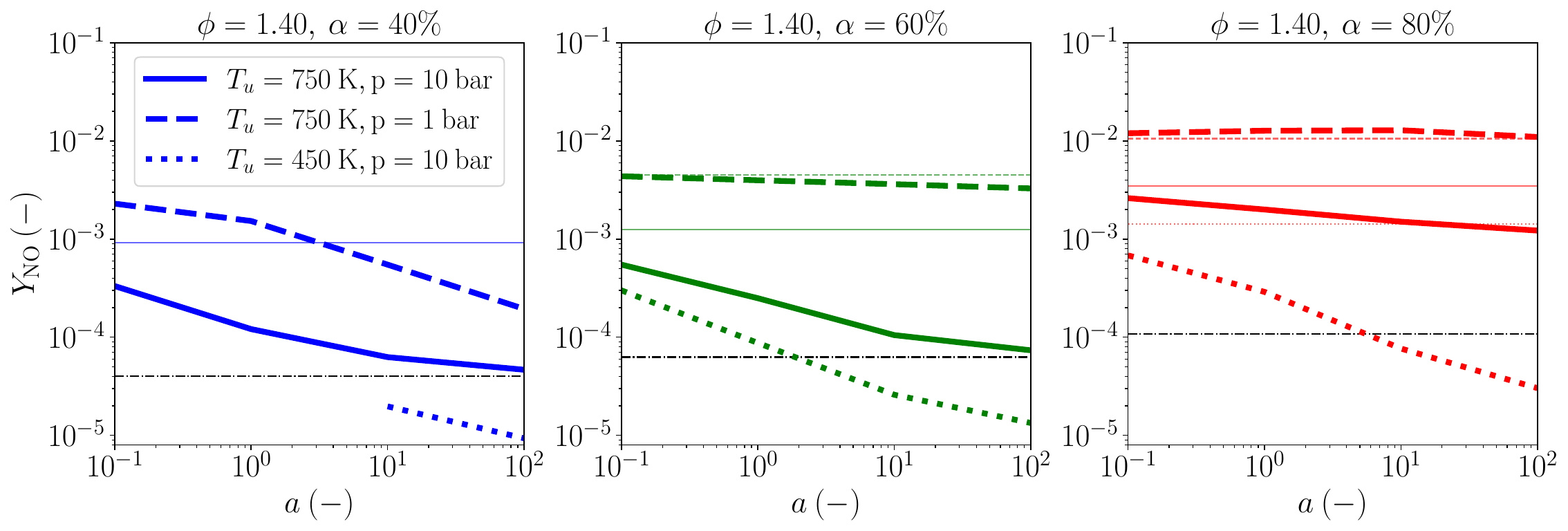}
\end{minipage}
\begin{minipage}{\textwidth}
  \centering
  \includegraphics[trim={0mm 0mm 0mm 10mm},clip,width=\textwidth]{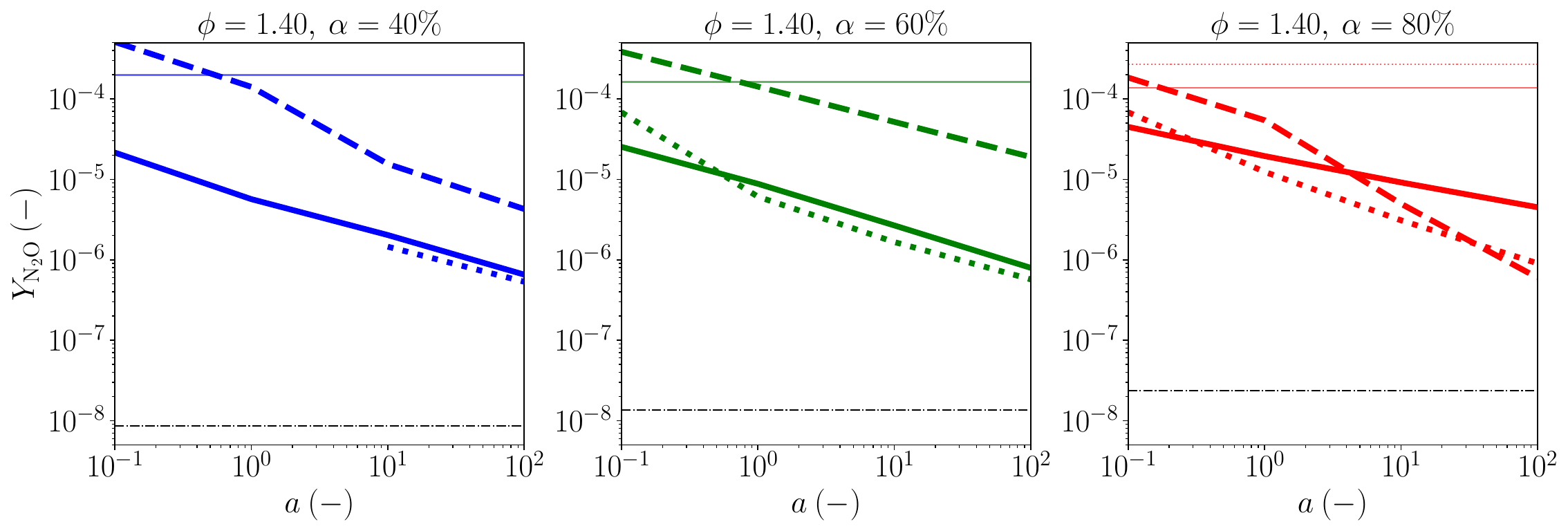}
\end{minipage}
\caption{Effect of pressure and temperature on reactive nitrogen emissions: $\NO$ (top) and $\NtwoO$ (bottom) mass fraction of the reacting flow particle at a 10 ms residence time as a function of the decay rate $a$, for $\chiLLnot=10^{4}\rm\:s^{-1}$. The thin transparent lines represent the same quantities but using $a=0$, and the black dash-dotted line represents thermodynamic equilibrium for the reference case (solid line).}
\label{fig:PT_YNO_YN2O_vs_a_end} 
\end{figure*}
The effects of pressure and initial temperature on reacting nitrogen production are now analyzed. In this case, the moderate mixing case $\chiLLnot=10^{4}\rm\:s^{-1}$, $p=10\rm\:bar$, and $T_{u}=750\rm\:K$ is kept as reference case. The second case considers lower pressure $p=1\rm\:bar$ and same initial temperature $T_{u}=750\rm\:K$. The third case considers lower initial temperature $T_{u}=450\rm\:K$ and same pressure. Figures~\ref{fig:PT_YNO_YN2O_vs_a_end} show $\NO$ (top) and $\NtwoO$ (bottom) emissions as a function of the decay rate $a$, for the three cracking percentages. The quantities for constant mixing ($a=0$, transparent lines) and equilibrium (for the reference case only, black line) are also included. In general, a decrease of pressure increases substantially the emissions, reaching over one order of magnitude difference for both $\NO$ and $\NtwoO$, with respect to the reference case. Trends of emissions with increasing decay rate changes from small to high cracking percentages, where $\NO$ goes from negative to zero slope and $\NtwoO$ towards more negative slopes. On the other hand, a decrease of initial temperature improves the reduction of reactive nitrogen, especially for moderate to high cracking percentages. For low cracking percentages, near equilibrium conditions are achieved for fast decays only. $\NO$ emissions are reduced by a factor of about two to five times with increasing decays, with respect to the reference case, which are even lower than the reference case equilibrium values. Although $\NtwoO$ emissions slightly decrease for small and moderate cracking, a reduction by a factor of four is observed for high cracking and very fast decay. Overall, high pressure and low initial temperature are convenient for low emissions. This could potentially benefit the upstream partial cracking processes.

Although it is meant to be used at the fuel-rich mixture stage or front-end of RQL combustors, the rush-to-equilibrium concept has also been evaluated for stoichiometric and lean conditions, which are included in the Supplementary Materials. It is shown that the concept still works in those conditions but with small to moderate reduction of reactive nitrogen emissions with respect to rich mixtures. It is also shown that the combustion chemistry model used utilized in simulations has no influence on the results.

\section{Practical implementation of the concept}
\label{sec:influence_TP}

The natural question that arises then is how the rush-to-equilibrium concept could be practically implemented. Assuming that the turbulent mixing rate scales with the turbulent velocity fluctuations that scale with the local mean velocity, the concept could be practically implemented within a diffuser-shaped combustion chamber. For a decay rate $a=1$ with an initial dissipation rate $\chi_{\Lambda\Lambda}^{0}=10^4\rm\:s^{-1}$, a typical combustor inlet diameter $D_i \sim\mathcal{O}(0.10\rm\:m)$~\cite{pratt1967performance, okafor2020control} and a diffuser angle $\theta=7^{\circ}$ \cite{lefebvre2010gas} would lead to an outlet diameter $D_o\sim\mathcal{O}(0.3\rm\:m)$ and combustor length $L\sim\mathcal{O}(0.9\rm\:m)$. A residence time of 10 $\rm ms$ would result in characteristics velocities of tens of $\rm m/s$. These quantities are reasonably consistent with gas turbine combustion chambers. A faster decay rate would require a larger diffuser angle (or a smaller diameter but longer combustor with a faster flow speed) so may not be practically relevant, but the decay rate $a=1$ already provides a substantial reduction in the reactive nitrogen emissions. The ideal decaying mixing rate that minimizes reactive nitrogen emissions to equilibrium conditions would require optimization of the combustion chamber geometry, which is left for future work.

\section{Conclusions} 
\label{sec:conclusions}

This work developed a novel concept to burn partially cracked ammonia at gas turbine conditions that enables low reactive nitrogen emission without increasing residence times. Reactive nitrogen emissions in AHN combustion are controlled by the very different approach to thermodynamic equilibrium compared to hydrocarbon fuels. In hydrocarbon combustion, reactive nitrogen emissions increase rapidly very near thermodynamic equilibrium favoring not too long residence times, but, in AHN combustion, reactive nitrogen emissions peak within the reaction zone and then relax toward thermodynamic equilibrium favoring long residence times. In AHN combustion, the ``unrelaxed'' emissions make up most of the reactive nitrogen emissions, so the rush-to-equilibrium combustion concept seeks to accelerate the approach to equilibrium without increasing residence time.

From fundamental budget analysis of premixed flames, while a fast mixing rate is required for transport within the preheat zone, fast mixing inhibits the approach to equilibrium. Therefore, in the rush-to-equilibrium combustion concept, a flow particle experiences a decaying mixing rate, faster in the preheat zone and decaying toward equilibrium. For typical gas turbine conditions, at a fixed residence time, the rush-to-equilibrium concept gets much closer to thermodynamic equilibrium. This results in substantially reduced reactive nitrogen emissions, on the order of an order of magnitude, for even modest mixing decay rates and even reaching thermodynamic equilibrium for faster decay rates. The rush-to-equilibrium concept could be practically implemented with a diffuser-shaped combustion chamber with very reasonable geometric and flow conditions for modest decay rates, and future work will focus on realization of the rush-to-equilibrium concept.

\section*{Acknowledgments}
\label{Acknowledgments}

The authors gratefully acknowledge funding from Princeton University through the Carbon Mitigation Initiative from the High Meadows Environmental Institute and the Energy Research Fund from the Andlinger Center for Energy and the Environment. The simulations presented in this article were performed on computational resources supported by the Princeton Institute for Computational Science and Engineering (PICSciE) and the Princeton University Office of Information Technology Research Computing department.

\bibliography{paper_RTE_CNF.bib} 
\bibliographystyle{elsarticle-num-names}

\end{document}